\documentclass[final,3p,times,authoryear]{elsarticle}

\usepackage{amssymb}
\usepackage{amsmath}

\usepackage{tabularx}
\usepackage{booktabs}
\usepackage{caption}
\usepackage{ragged2e}

\usepackage{hyperref}
\usepackage{makecell}

\usepackage[figuresright]{rotating}
\usepackage[section]{placeins}

\usepackage{verbatim}
\usepackage{color}

\usepackage{etoolbox}
\makeatletter
\def\ps@pprintTitle{%
  \let\@oddhead\@empty
  \let\@evenhead\@empty
  \def\@oddfoot{\centerline{\thepage}}%
  \let\@evenfoot\@oddfoot}
\makeatother

\begin{document}

\begin{frontmatter}



\title{Paint by Odor: An Exploration of Odor Visualization through Large Language Model and Generative AI} 


\affiliation[label1]{
            organization={Academy of Arts and Design, Tsinghua Unviersity},
            city={Beijing},
            postcode={100084},
            country={China}}

\affiliation[label2]{
            organization={The Future Laboratory, Tsinghua University},
            city={Beijing},
            postcode={100084},
            country={China}}

\affiliation[label3]{
            organization={School of Architecture, Tsinghua University},
            city={Beijing},
            postcode={100084},
            country={China}}

\affiliation[label4]{
            organization={School of Digital Media and Design Arts, Beijing University of Posts and Telecommunications},
            city={Beijing},
            postcode={100084},
            country={China}}

\author[label1,label2]{Gang Yu}
\author[label1,label2]{Yuchi Sun}
\author[label2,label4]{Weining Yan}
\author[label2,label3]{Xinyu Wang}
\author[label2]{Qi Lu\corref{corresponding}}

\cortext[corresponding]{Corresponding author (luq@mail.tsinghua.edu.cn).}

\begin{abstract}
Odor visualization translates odor information and perception into visual outcomes and arouses the corresponding olfactory synesthesia, surpassing the spatial limitation that odors can only be perceived where they are present. Traditional odor visualization has typically relied on unidimensional mappings, such as odor-to-color associations, and has required extensive manual design efforts. However, the advent of generative AI (Gen AI) and large language models (LLMs) presents a new opportunity for automatic odor visualization. Nonetheless, gaps remain in bridging olfactory perception with generative tools to produce odor images. To address these gaps, this paper introduces \textit{Paint by Odor}, a pipeline that leverages Gen AI and LLMs to transform olfactory perceptions into rich, aesthetically engaging visual representations. Two experiments were conducted, where 30 participants smelled real-world odors and provided descriptive data and 28 participants evaluated 560 generated odor images through seven systematically designed prompts. Our findings explored the capability of LLMs in producing olfactory perception by comparing it with human responses and revealed the underlying mechanisms and effects of language-based descriptions and several abstraction styles on odor visualization. Our work further discussed the possibility of automatic odor visualization without human participation. These explorations and results have bridged the research gap in odor visualization using LLMs and Gen AI, offering valuable design insights and various possibilities for future applications.
\end{abstract}




\end{frontmatter}


\section{Introduction}
\label{sec1}
Odors play a critical role in many situations, such as during environmental hazard avoidance (\cite{pence_risk_2014}), nutritional analysis (\cite{boesveldt_importance_2021}), social relationship regulation (\cite{ravreby_there_2022, mishor2021sniffing}), and elicitation of emotions and memories (\cite{lu_exploring_2020, amores_fernandez_olfactory_2023, buettner_aversive_2017}). However, due to the chemical nature of smells, olfactory perception is limited to the physical spaces within which odorants are present. One approach to expressing olfactory perception beyond spatial constraints is to utilize \textit{olfactory imagery}, which Richard and Trevor defined as the ability to experience the sensation of smell in the absence of an appropriate stimulus (\cite{stevenson2005olfactory}). While olfactory imagery predominantly describes a participant's cognitive ability to evoke mental impressions of smells, this paper focuses on a more specific scope: \textit{odor visualization}, referring to the visual outcomes inspired by olfactory experiences. For instance, many perfume brands have successfully associated their products with particular colors or color schemes (\cite{spence_olfactory-colour_2020}). A well-known example is the iconic gold and yellow theme for Chanel N\textsuperscript{o}5. In addition to commercial applications, odor visualization has also been employed in art installations (\cite{ueda_website, hanahana2010, Cao2015Pebbles}) to enhance multisensory interactive experiences. These efforts to translate olfactory sensations into thoughtfully designed visuals have opened avenues for enabling remote and digital olfactory experiences.

However, such works have generally relied on uni-dimensional or bi-dimensional mappings, such as odor-to-color or odor-to-shape associations, which cannot adequately capture the complexity of human olfaction. Since olfactory perception encompasses multidimensional aspects (sensory, emotional, mnemonic, and cognitive), we propose that by generating more feature rich, integrated images, we are able to better express olfactory information and its accompanying feelings. Moreover, existing odor visualizations are often created on a case-by-case basis, with few computer-aided or automated systems specifically designed for odor visualization, in contrast to other fields (\cite{lin_autoposter_2023, liu_opal_2022}). Nevertheless, the emergence of generative AI (Gen AI) and large language models (LLMs) introduces a new avenue for automatic odor visualization via language descriptions, enabling AI models to produce olfactory perception \cite{zhong2024sniff} and generate corresponding odor images. Although the effectiveness of language-based odor descriptors is subject to debate (\cite{olofsson_olfactory_2021, croijmans2015odor, olofsson_muted_2015}), they remain the most feasible method to define, quantify, and represent olfactory perceptions systematically (\cite{keller_olfactory_2016, lee_principal_2023, dravnieks1992atlas}). Using prompt engineering, one can embed rich olfactory qualities within image content. Nonetheless, key questions still remain: to what extent can LLMs produce human-aligned olfactory perception from a certain odor, and what principles enable effective prompting for generating accurate odor images by Gen AI?

To address these gaps, we present \textit{Paint by Odor}, a pipeline for odor visualization that integrates LLMs and Gen AI to transform olfactory perceptions into rich, aesthetically compelling images. In this pipeline, structured odor descriptions are generated by both users and an LLM. A specialized LLM, trained on cross-modal research, then translates these odor descriptions into graphical cues. Building on these, we design seven prompts derived from different combinations of descriptive and stylistic elements, each reflecting principles for odor visualization. To deepen our understanding of LLM-based olfactory perception, examine user cross-modal synesthesia, glean insights and guidelines for more effective prompt design, investigate user preferences, and envision future applications, we conducted two experiments. In the first, participants smelled 20 common odors and selected descriptors reflecting both olfactory and emotional perceptions, while the LLM was also tasked with generating descriptors for the same odors. Comparison revealed that the LLM offered a broader but somewhat stereotyped perspective. In the second study, we used both sets of descriptors to generate odor images, which participants evaluated through quantitative questionnaires and qualitative interviews. 
Thematic analyses then highlighted findings in odor visualization.

To the best of our knowledge, this study is the first attempt to use LLM and language-based descriptions to produce complex images for conveying olfactory information. By answering our research questions, we provide the following contributions:

\begin{enumerate}
\item 	Development of \textit{Paint by Odor}, a pipeline that utilizes LLMs and Gen AI to generate richly detailed, aesthetically engaging odor-semantic images.
\item	A sensory study of 20 household odors to evaluate human olfactory perceptions alongside LLM-generated interpretations of the same olfactory stimuli.
\item	Qualitative and quantitative analysis of key visual elements, the role of different descriptors, and stylistic comparisons, culminating in design guidelines for odor visualization.
\item	Insights into future applications of odor visualization, positioning our system as both a research tool for studying visual-olfactory crossmodal interactions and a design tool for commercial and artistic use.
\end{enumerate}

\section{Related Work}
\label{sec2}
\subsection{Odor Visualization}

The visualization of odors has been explored in marketing and advertisement. Krishna et al. found the necessity of olfactory imagery on consumers' physiological, evaluative, and consumptive responses to advertised food products (\cite{krishna2014smellizing}). Varun and Zachary investigated the largest retailer in the US and found that 27\% of scented laundry detergents and cleaners included a picture of the scented object on their packages and further demonstrated a conceptual model for systematically understanding how vision can be utilized to evoke the benefits of scent for consumer behavior (\cite{sharma2024seeing}). Similar effects of olfactory imagery or odor visualization also present in perfumery (\cite{velasco2006olfactory}). Apart from commercial products, odor visualization is applied to interactive art installations. In the \textit{Digital Provence} project, TeamLab and L’Occitane created the ambiance of Provence by integrating visual and olfactory experiences. As visitors approached a fragrance wall, the system emitted floral scents and projected virtual flowers onto a mirrored wall (\cite{Lai2017Scent}). \textit{Hanahana} was designed as an interactive vase utilizing scent data as input and generating visual output to display the existence and variation of ambient scents (\cite{hanahana2007, hanahana2010}). Cao et, al. proposed \textit{Scented Pebbles}, an interactive installation integrating multi-sensory ambience of light and smell (\cite{Cao2015Pebbles}). These works usually present visual forms and smell at the same time, and create the visual experience from the sense of smell, which can be called ``We see what we smell'' or ``Vision-O-Smell'' (\cite{lai_scent_2017}). Another related field is urban odor visualization, exemplified by Kate McLean’s olfactory mapping of Amsterdam (\cite{mclean2017smellmap}). Kate’s odor maps, informed by participants’ self-reported perceptions, used visual elements to depict historical and locational smells, effectively conveying olfactory experiences across spatial and temporal boundaries (\cite{mclean2017smellmap, mclean2019nose}). These maps, typically featuring color areas overlaid on city maps, represent semantic information about urban odors, including their sources, intensities, and trajectories.

Odor visualizations on product packaging and advertising materials are typically composed of singular, concrete objects. For olfactory art installations, simple colors, lines, lights, and other graphical elements are typically used. Although such elements successfully convey semantic scent-related information and create a multisensory experience, these visualizations inadequately capture the rich and multifaceted features of olfactory perception, which presents a market gap and opportunity for employing Gen AI to create more complex and aesthetic visualizations.

\subsection{Visual-olfactory cross-modal correspondence}
The relationship between visual and olfactory modalities has been extensively researched in many different fields including psychology, physiology, neuroscience, and design. It is one of the synaesthesis pairs of sensory modalities that describes an olfactory (or visual) sensory feature to be associated with a visual (or olfactory) sensory feature (\cite{spence_cognitive_2012, simpson_types_1955}). Among various visual features, color-odor correspondence has emerged as the most thoroughly researched topic. Early studies demonstrated that mismatched or masked colors could hinder the identification of odors or flavors (\cite{r_j_stevenson_effect_2008, dubose_effects_1980, morrot_color_2001}). Subsequent research has explored the effects of color properties such as hue, lightness, and saturation on color-odor matching, gradually uncovering the systematic relationships between these features (\cite{spence_olfactory-colour_2020, kim_can_2013}). These correspondences have been reported to exhibit stable patterns (\cite{avery_n_gilbert_cross-modal_1996, luisa_dematte_cross-modal_2006}), both in terms of consistencies and variations across cultures (\cite{spence_olfactory-colour_2020, nehme_influence_2016, jacquot_colours_2016, levitan_cross-cultural_2014}). Beyond color, studies have investigated correspondences between odors and visual shapes or textures (\cite{ryan_j_ward_smelling_2022, g_hanson-vaux_smelling_2013, kaeppler_crossmodal_2018, m_luisa_dematte_cross-modal_2006}). There are several ways to explain the mechanisms behind them, such as associations of odor objects (\cite{kaeppler_crossmodal_2018, speed_odorcolor_2023, goubet_seeing_2018}), learning and natural biases (\cite{spector_making_2012}), perceptual attributes of odors (\cite{stevenson_nature_2012}), and emotions (\cite{schifferstein_visualising_2004}). Recent advances in neuroscience have also examined the neural linkages between visual and olfactory modalities, shedding light on the underlying mechanisms (\cite{rolls_orbitofrontal_1996, jay_a_gottfried_nose_2003, sijben_semantic_2018, jadauji_modulation_2012, kehl2024single}). Charles et al. categorized these mechanisms into four main groups: internalized environmental statistics, emotional mediation, shared neural representations, and the use of common lexical terms (\cite{spence_olfactory-colour_2020}). Collectively, these studies provide essential theoretical foundations for odor visualization and its synesthetic perception.

However, much of the existing research focuses on isolated visual elements, neglecting the holistic experience of viewing integrated images. This limitation undermines the potential of evoking richer olfactory synesthesia. In our work, we aim to bridge this gap by integrating multiple visual elements into cohesive images and evaluating their capacity to elicit olfactory synesthesia.

\subsection{Description for Odors}
Among human sensory experiences, olfaction is regarded as evocative but elusive due to the difficulty of describing it (\cite{iatropoulos_language_2018}). Some research have reported that our ability to name the source of odors is exceptionally unimpressive (\cite{olofsson_olfactory_2021, croijmans2015odor, olofsson_muted_2015}) and it might be attributed to the loss of signal fidelity across neural olfactory-language processing stages (\cite{nehme_influence_2016}). Despite this, some studies suggest that basic odor terms are universally shared across languages and that certain cultures excel in naming odors and discussing olfactory experiences (\cite{majid_olfactory_2018, majid2021human}). To solve the challenges in describing olfactory experiences, researchers have developed standardized odor descriptor systems. Dravnieks, for instance, conducted sensory experiments to construct a dataset comprising 160 odorants and 146 descriptors (\cite{dravnieks1992atlas}). Similar datasets include Keller’s 480 odorants with 20 odor descriptors (\cite{keller_olfactory_2016}), Ma’s 222 binary mixture odorants with 2 dimensions, namely odor intensity and pleasantness, and GS-LF with 138 odor descriptors for nearly 5000 molecules (\cite{lee_principal_2023}). These datasets were created through psychophysical experiments where participants rated odors against predefined descriptors. The descriptor-based olfactory language has formed the foundation for evaluating anosmia \cite{croy_peripheral_2015}), predicting olfactory perceptions (\cite{lee_principal_2023, keller2017predicting}), and constructing odor spaces (\cite{lee_principal_2023, ravia2020measure}).

However, most of these datasets focus on single molecules with explicit chemical structures, not considering real-world scenarios where odors typically consist of complex mixtures. To address this gap, our work adapts these methodologies to develop a new perceptual dataset based on real-world odor sources. This dataset serves as the foundational data for our design of odor visualization.

\section{Design Principles and System Framework}
\label{sec3}
\subsection{Design Principles}
To guide our odor visualization pipeline, we first examined the principles underlying odor visualization. The authors conducted extensive literature reviews in olfactory science, cognitive psychology, HCI, graphic design, and generative AI, followed by several rounds of discussions with HCI researchers, visual designers, and olfactory experts.

These discussions yielded design principles from the following dimensions: (1) Physical properties of odors, (2) Emotions and memories, (3) Complexity of odorant molecules and their multidimensional mappings, (4) Methods for quantifying and describing olfactory experiences, (5) Technical advantages of Gen AI and LLMs, (6) Graphic design basics (compositional balance, visual hierarchy, color theory), (7) Visual aesthetics for translating olfactory experiences, and (8) Future applications. From this, we formulated three Gen-AI-based principles for scent visualization.

\begin{enumerate}[$\bullet$]
\item \textbf{Principle 1: Multi-sensory Mapping with Rich Graphic Elements.} Traditional single-dimensional mappings (e.g., odor-to-color or odor-to-shape) cannot capture the multifaceted nature of human olfactory perception, which often involves physiology, object association, emotions, and memories. In addition, odors frequently appear as mixtures, leading to complex smell perception. Thus, a single image should encode multiple olfactory sensations. Since imagination and memories are highly personalized, we emphasize physical and emotional properties as potentially shared aspects of olfactory perception between people. Appropriate mappings reinforce odor visualization, enabling precise conveyance of olfactory perception. Our system integrates diverse visual elements to generate holistic images encoding these varied sensations.

\item \textbf{Principle 2: Language-based Description as Mediation between Olfaction and Gen AI.}
Although the adequacy of language to describe olfactory perceptions is debated, odor descriptors remain widely used to systematically define, quantify, and represent smell experiences. Given that modern Gen AI and LLMs rely on textual prompts, language is the most feasible mediator in our odor visualization pipeline. People describe olfactory sensations using physical and emotional descriptors, while the LLMs process this text, mapping key imagery features and translating odor descriptions into prompts for image generation. By connecting these two linguistic dimensions, our system provides a smooth pathway from smell descriptions to visual representation.

\item \textbf{Principle 3: Hierarchical Visual Structure and Aesthetic Viewing Experience.}
Odor visualization spans product design to artistic installations, requiring high aesthetic quality. By decomposing the visual composition, we extracted five fundamental imagery elements — color, shape, line, texture, and style — from odor perceptions. These elements should work collectively to convey olfactory characteristics while providing artistic value. We adopt abstract painting as our primary approach because (1) Odors are often invisible and colorless, making abstraction suitable; (2) The non-representational character of abstract art aligns with the diffuse nature of olfactory perception; (3) Abstraction can harmonize diverse elements into cohesive compositions. We designed our pipeline and experiments to assess this hypothesis.
\end{enumerate}

\subsection{Paint by Odor: Framework}

\subsubsection{Overview}

\begin{figure}[t]
\centering
\includegraphics[width=0.9\columnwidth]{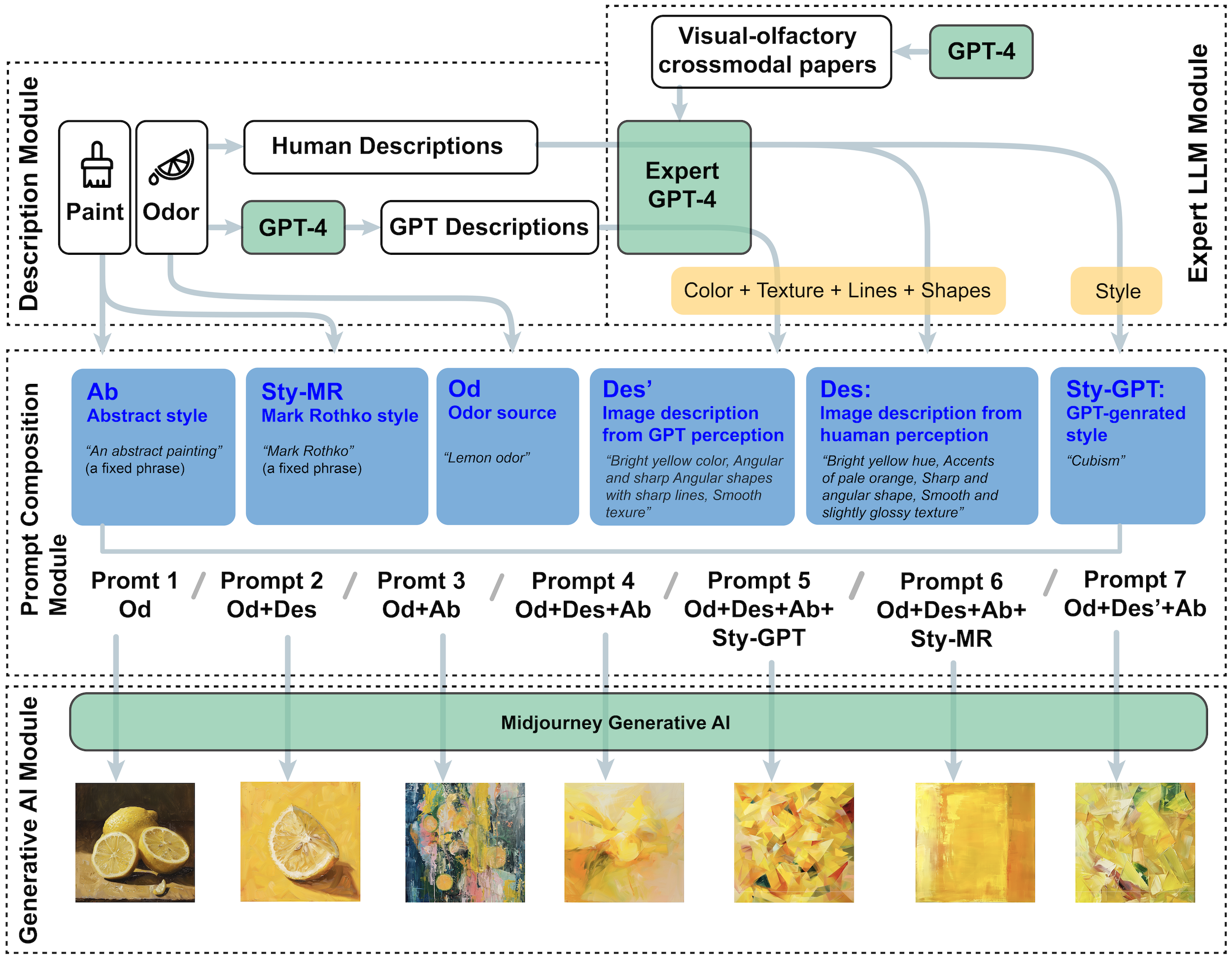}
\caption{The system schematic of \textit{Paint by Odor}, a pipeline that utilizes LLMs and Gen AI to generate odor images.}\label{fig:system_framework}
\end{figure}

Guided by these design principles, we developed Paint by Odor, a pipeline leveraging LLMs and Gen AI to visualize human olfactory perception via rich graphical elements. The pipeline comprises of four main modules: (1) a description module, (2) an expert LLM module, (3) a prompt composition module, and (4) a Gen AI module. Figure~\ref{fig:system_framework} illustrates our framework.

\subsubsection{Odor Description Unit}
Our system starts with a specific odor. We designed two strategies to obtain odor descriptions: one from human perception, and another from a large language model (GPT-4). These descriptions combine odor (e.g. `fruity,' `woody') and emotional (e.g. `calming,' `hopeful') descriptors, capturing both material and emotional dimensions. This foundational step follows findings from a preliminary study described in Chapter~\ref{sec4}.

\subsubsection{Expert GPT Unit}
Because commercial generative AI systems are not specifically optimized for visualizing olfactory perceptions, we developed an expert GPT fine-tuned on 20 visual-olfactory crossmodal papers (see Table~\ref{tabS2} in Appendix). This unit specializes in translating odor descriptions into imagery descriptions, focusing on colors, textures, and lines \& shapes. A fourth element, style, provides an additional layer of artistic interpretation.

\subsubsection{Prompt Unit}
\begin{figure}[t]
    \centering
    \includegraphics[width=0.9\columnwidth]{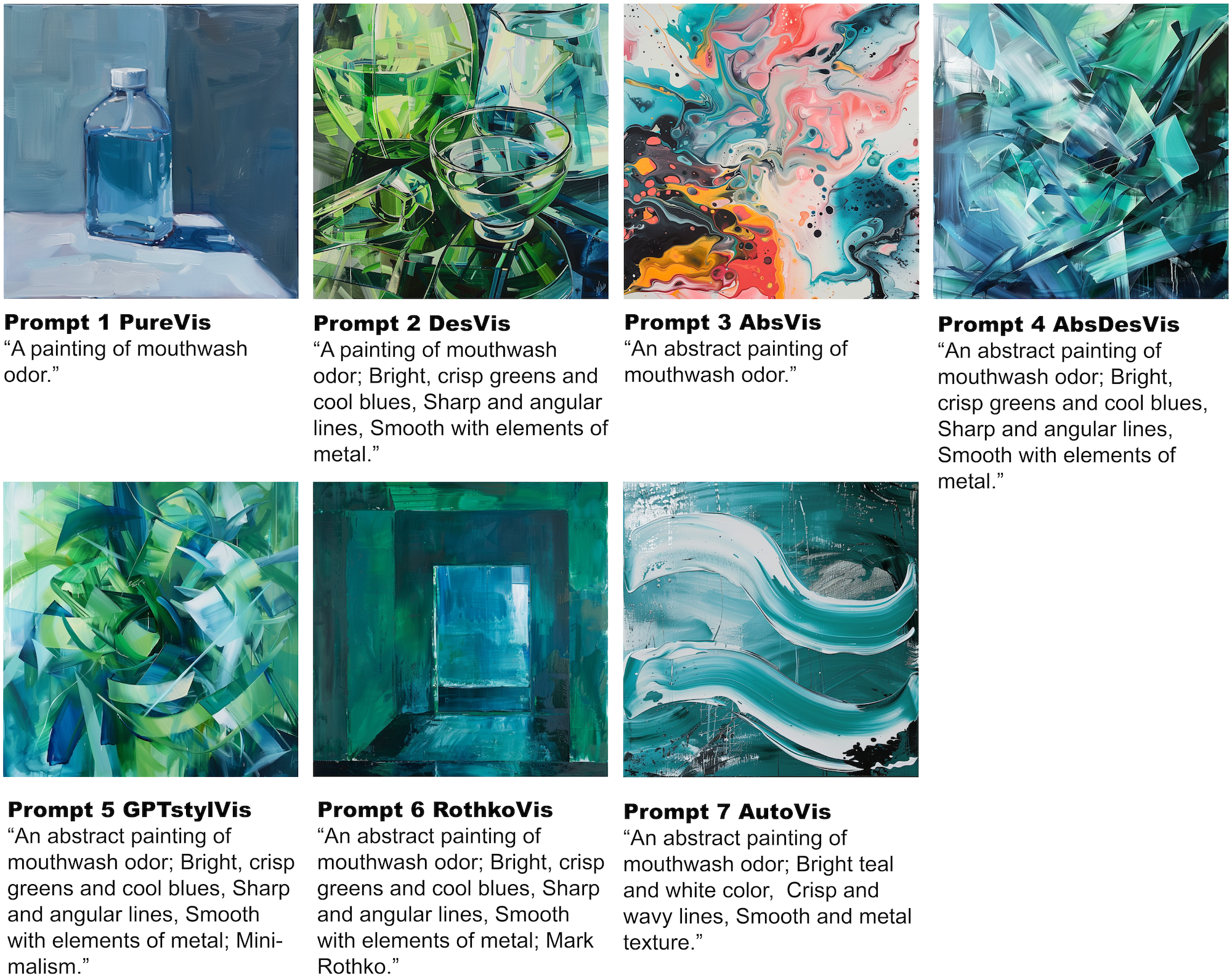}
    \caption{Image examples of Mouthwash smell and corresponding prompts. Each image corresponds to one of seven prompt inputs to Midjourney.}\label{fig:prompt_example}
\end{figure}

In line with our three principles, we crafted prompt elements and prompts that capture odor sources, descriptions, and stylization. In total, six prompt elements and seven different prompts were designed, each representing a distinct path to generate odor images. Figure~\ref{fig:prompt_example} shows an example for Mouthwash. The six elements are introduced below:

\begin{enumerate}[$\bullet$]
    \item \textbf{\textit{Od}:} This unit focuses on describing the odor source, such as lemon. The prompt phrase is the name of substances.
    \item \textbf{\textit{Des}:} We incorporate image descriptions generated by the expert GPT-4 model. This unit originates from human perceptions collected in the preliminary study and emphasizes three key dimensions: color, texture, and lines\&shapes. We asked GPT for commonly used textures and formed a texture set for Expert GPT to choose from, including impasto, rough, collage, smooth, corrugated, engraving, fabric, metal, and frottage.
    \item \textbf{\textit{Des}':} Similar to Prompt Des but based on LLM-generated (an original GPT-4) descriptions of the odor.
    \item \textbf{\textit{Abs}:} This unit specifies an abstract painting style with the fixed phrase ``An abstract painting''.
    \item \textbf{\textit{Sty-MR}:} This unit emphasizes the style of Mark Rothko, a famous artist known for his color-composed abstract paintings which evoke human emotions. We chose this style  after a few steps of filtering from Midlibrary (\cite{midlibrary_website}), a style collection for Midjourney. We first browsed the ``abstract styles'' class and sorted out 75 styles that generated stable painting effects. Most of these styles were named by their artists, while some were named by specific art schools. Then we grouped them into four categories, and chose one representative style for each group. including Encaustic paint or hot wax painting (for fluidic styles), Sam Francis's style (for splash-ink styles), Mark Rothko's style (for color-cubic styles), and Emily Kame Kngwarreye's style (for abundant color styles). To simplify our experiment, five authors discussed and voted to choose Mark Rothko's as the final customized abstract style. Rothko's works are often seen as abstractions of human feelings, preserving aesthetic judgment. We choose this style to enrich the olfactory and emotional expressions. The prompt unit refers to a library called Midlibrary, specifically tailored for the Midjourney platform.
    \item \textbf{\textit{Sty-GPT}:} This unit allows the expert GPT to propose a suitable abstract style based on the odor description, offering flexibility and alternative interpretations. Similar to textures, we first asked GPT for common art schools and created a set of styles for the expert GPT to choose from. The styles included cubism, impressionism, surrealism, expressionism, minimalism, futurism, dadaism, art nouveau, pop art, and constructivism.
\end{enumerate}

Based on the above six elements, seven different prompts for odor visualization were created, as shown in Figure \ref{fig:prompt_example} with examples for Mouthwash.

\begin{enumerate}[$\bullet$]
    \item \textbf{Prompt 1 \textit{PureVis} (Pure Visualization): Od} This prompt was designed to illustrate concrete paintings of odors.
    \item \textbf{Prompt 2 \textit{DesVis} (Description Visualization): Od + Des} This prompt was designed to encode human olfactory and emotional perception into concrete paintings.
    \item \textbf{Prompt 3 \textit{AbsVis} (Abstract Visualization): Od + Abs} This prompt was abstract style paintings of odors without any perceptual descriptors.
    \item \textbf{Prompt 4 \textit{AbsDesVis} (Abstract Description Visualization): Od + Des + Abs} This prompt was designed to visualize human perceptual information into abstract paintings.
    \item \textbf{Prompt 5 \textit{GPTstylVis} (GPT-style Visualization): Od + Des + Abs + Sty-GPT} This prompt encoded a specific abstract style extracted by the expert GPT, others remained the same as Prompt 4.
    \item \textbf{Prompt 6 \textit{RothkoVis} (Rothko's style Visualization): Od + Des + Abs + Sty-MR} This prompt adopted Mark Rothko’s style for a special visualization style.
    \item \textbf{Prompt 7 \textit{AutoVis} (Automated Visualization): Od + Des’ + Abs} This prompt represented the automatic generation of our system without human perception and description.
\end{enumerate}

\subsubsection{Generative AI Unit}

We selected Midjourney, a leading generative AI model, to produce visuals from our seven prompts. Implementation details and evaluation findings are in Chapter~\ref{sec5}.

\subsection{Research Questions}
As the pioneering effort to apply Large Language Models (LLM) and generative AI to odor visualization, this paper encountered several unexplored questions. Although the ability of AI recognizing scents from human description has been explored \cite{zhong2024sniff}, the capability boundaries of LLM and Gen AI in interpreting odors and producing olfactory perception still remain unknown. Additionally, the optimal methods for designing prompts to transform olfactory perceptions into visual representations require further investigation and evaluation. It is also valuable to explore participant preferences and application scenarios for odor images, considering both semantic correspondence and aesthetic experience. To explore and bridge these gaps, we propose several research questions and designed two experiments. The research questions are as follows:

\textbf{RQ1} How capable are large language models of perceiving and understanding odor?

\textbf{RQ2} How do rich graphical elements influence participants’ olfactory synesthesia, and what are the underlying mechanisms?

\textbf{RQ3} What is the mediating effect of language-based description in transforming olfactory experiences into images?

\textbf{RQ4} How do artistic styles, particularly abstraction, impact scent portrayal and image aesthetics? What are users’ style preferences?

\section{Preliminary Study: Description on Odors, Human VS LLM}
\label{sec4}
In this study, participants smelled 20 common odors and selected descriptors from a prepared list of odor or emotion descriptors. Concurrently, we queried GPT-4 with the same questions to generate descriptor lists. We then compared GPT’s perception to human perception, addressing \textbf{RQ1}.

\subsection{Method}
\subsubsection{Odor preparation}
We chose 20 common household odors spanning food, beverages, fragrances, cleaning products, and daily necessities. All odors are safe because they commonly occur in daily life. We used two types of containers: a wide-mouth glass bottle for moderate odorants and a small reagent bottle for pungent ones. Table~\ref{tab1} lists each odor and its container type.

\begin{table}[h]
\begin{tabular}{@{}lll@{}}
\toprule
\textbf{Odor Source}        & \textbf{Bottle Type} & \textbf{Status}          \\ \midrule
Lemon                       & wide-mouth bottle    & fresh fruit, cut in half \\
Grape                       & wide-mouth bottle    & fresh fruit, cut in half \\
Green Tea                   & wide-mouth bottle    & solid fragments          \\
Baijiu                      & small reagent bottle & liquid                   \\
Coffee Beans                & wide-mouth bottle    & solid                    \\
Vinegar                     & small reagent bottle & liquid                   \\
Sesame Oil                  & small reagent bottle & liquid                   \\
Unburnt Cigarette (Tobacco) & wide-mouth bottle    & separated from cigarette \\
Lily Flower                 & wide-mouth bottle    & original lily petals     \\
Latex Paint                 & small reagent bottle & liquid                   \\
Varnish                     & small reagent bottle & liquid                   \\
Mark Pen Ink                & small reagent bottle & liquid                   \\
Glue                        & small reagent bottle & liquid                   \\
Shampoo                     & wide-mouth bottle    & liquid                   \\
Shower Gel                  & wide-mouth bottle    & liquid                   \\
Detergent                   & small reagent bottle & liquid                   \\
Mouthwash                   & small reagent bottle & liquid                   \\
Osmanthus Essential Oil     & small reagent bottle & liquid                   \\
Shoe Polish                 & small reagent bottle & liquid                   \\
Laundry Liquid              & wide-mouth bottle    & liquid                   \\ \bottomrule
\end{tabular}
\caption{Odors chosen for the preliminary study and their containers. Preparation status of the odor substance is listed, including existing phases and methods of pretreatment.}
\label{tab1}
\end{table}

\subsubsection{Odor descriptors}
The odor descriptor set originated from the GS-LF descriptors (\cite{lee_principal_2023}), initially 138 terms. To minimize experimental fatigue and enhance clarity/understandability, we leveraged six olfactory researchers to remove near-duplicates and less relevant terms. Recognizing the lack of descriptors for malodorous odorants, paint samples, and oil samples, we added `ozone', `oily', and `waxy' to the word list. In total, 34 odor descriptors suited to our odor samples were ultimately derived. 

To evaluate emotional responses, we adopted a six-pair Semantic Differential Scale from Mehrabian and Russell (\cite{ray_chatgpt_2023, russell1974approach}), adjusted for olfactory context. We added `admire-disgust,' referencing prior olfactory emotion research (\cite{ekman_argument_1992, croy_basic_2011, bestgen_odor_2015, tang_admired_2022}). Both odor and emotion descriptors were presented in English and Chinese.

\subsubsection{Participant}
A total of 30 participants were recruited for this study, including 13 males (43.3\%) and 17 females (56.7\%) with ages ranging from 20 to 49 (mean = 25.83, std = 5.73). Prior to the experiment, all participants confirmed that they were not experiencing any symptoms of cold, nasal congestion, or stuffiness that could potentially affect their olfactory perception. Each participant received CNY 60 and the experiment was approved by the Tsinghua University Science and Technology Ethics Committee (Medicine).

\subsubsection{Procedure}

\begin{figure}[htb]
\centering
\includegraphics[width=1\columnwidth]{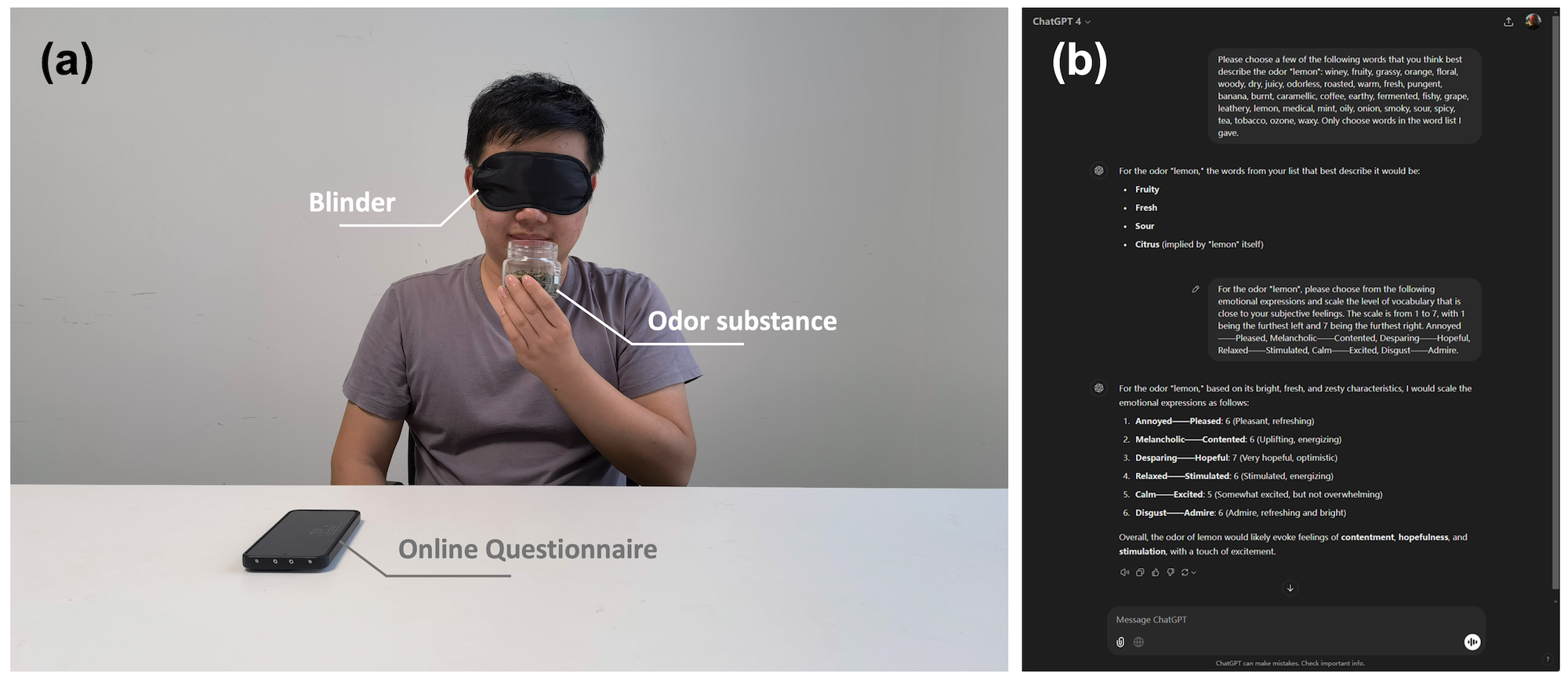}
\caption{(a) Picture of the preliminary study. Participants were blindfolded while sniffing odors, and subsequently filling out the questionnaire using the mobile phone. (b) GPT's olfactory perception. Researchers asked GPT the same questions in the questionnaire including choosing odor descriptors and scoring emotions.}
\label{fig:user_study_1_pic}
\end{figure}

The experiment took place in a 3m x 3m office (Figure~\ref{fig:user_study_1_pic}(a)). Since visual cues about odor sources can influence olfactory perception (\cite{r_j_stevenson_effect_2008, dubose_effects_1980, morrot_color_2001}), we required participants to wear blindfolds while sniffing. After smelling, participants completed a questionnaire choosing appropriate odor descriptors and rating emotional responses on 7-point Likert scales. Following the completion of the questionnaire for each odorant, participants were encouraged to share any additional feelings or comments related to their olfactory perception.

For LLM capabilities, we queried GPT-4 (April 2024 version) with the same questions, asking it to pick odor/emotion descriptors from the given list. We also asked GPT for clarification on the reasons for choosing certain odor and emotion descriptors. Figure~\ref{fig:user_study_1_pic}(b) shows the conversational interface.

\subsection{Prelminary Findings}

\begin{figure}[htb]
\centering
\includegraphics[width=1\columnwidth]{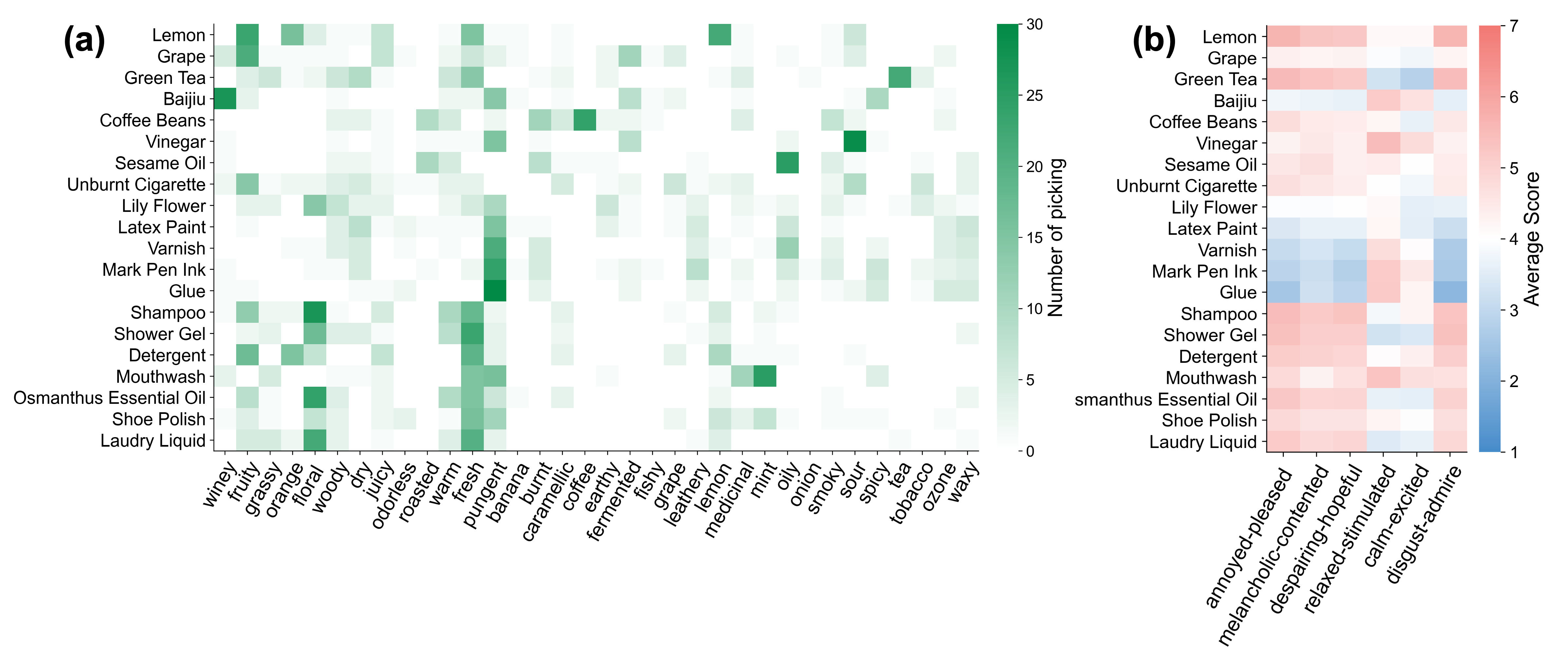}
\caption{Results of questionnaires. (a) Results of odor descriptors picked by participants. The numbers came from the sum of participants who chose a particular descriptor for a particular odor. The deeper colors are, the more participants chose them. (b) Participants' average scores of emotional descriptors. The deep blue color meant negative emotions (score 1\textasciitilde3), while the deep red color meant positive emotions (score 5\textasciitilde7). The light gray area represented neutral emotions (score 3\textasciitilde5).}
\label{fig:user_study_1_results}
\end{figure}

\begin{sidewaystable}[htb]
\resizebox{\textwidth}{!}{%
\begin{tabular}{@{}l|ll|ll@{}}
\toprule
\textbf{Odor Substance} & \textbf{Odor Descriptor} & \textbf{Emotion Descriptor} & \textbf{Odor Descriptor (GPT)} & \textbf{Emotion Descriptor (GPT)} \\
\midrule
Lemon & fruity, orange, lemon & \makecell[tl]{pleased, contented, \\ hopeful, admire}  & \makecell[tl]{fruity, fresh, citrus, \\ lemon} & \makecell[tl]{pleased, contented, stimulated, \\ excited, admire} \\
Grape & fruity, juicy, fermented & - & fruity, juicy, fresh, grape & \makecell[tl]{pleased, contented, stimulated, \\ excited, admire} \\
Green Tea & dry, fresh, tea & \makecell[tl]{pleased, contented, \\ hopeful, calm, admire} & \makecell[tl]{earthy, floral, woody, \\ fresh, dry, roasted, mint} & \makecell[tl]{pleased, contented, hopeful, \\ relaxed, admire} \\
Baijiu & winey, pungent, spicy & stimulated & \makecell[tl]{winey, fruity, earthy, \\ fermented, pungent} & stimulated, excited, admire \\
Coffee Beans & roasted, burnt, coffee & - & \makecell[tl]{roasted, warm, earthy, \\ caramellic, coffee, burnt} & \makecell[tl]{pleased, contented, stimulated, \\ excited, admire} \\
Vinegar & pungent, fermented, sour & stimulated & \makecell[tl]{sour, pungent, fermented, \\ acrid} & stimulated, admire, disgust \\
Sesame Oil & roasted, burnt, oily & - & roasted, warm, nutty, oily & pleased, contented, admire \\
\makecell[tl]{Unburnt Cigarette \\ (Tobacco)} & fruity, grape, sour, tobacco & - & \makecell[tl]{tobacco, earthy, woody, \\ leathery} & relaxed, disgust, admire \\
Lily Flower & floral, pungent, earthy & - & floral, fresh, sweet & pleased, contented, admire \\
Latex Paint & dry, pungent, oily, waxy & - & chemical, pungent, fresh & stimulated, disgust, excited \\
Varnish & \makecell[tl]{dry, pungent, \\ burnt, oily, waxy} & disgust & pungent, chemical, woody & stimulated, disgust, excited \\
Mark Pen Ink & pungent, leathery, spicy & \makecell[tl]{annoyed, despairing, \\ stimulated, disgust} & chemical, pungent, sharp & stimulated, disgust, annoyed \\
Glue & \makecell[tl]{pungent, burnt, \\ ozone, waxy} & \makecell[tl]{annoyed, despairing, \\ stimulated, disgust}  & chemical, pungent, sharp & stimulated, disgust, annoyed \\
Shampoo & fruity, fresh, pungent & \makecell[tl]{pleased, contented, \\ hopeful, admire} & fruity, floral, fresh, mint & pleased, contented, excited \\
Shower Gel & floral, warm, fresh & \makecell[tl]{pleased, contented, \\ hopeful, admire} & fruity, floral, fresh, mint & pleased, contented, stimulated \\
Detergent & fruity, orange, fresh & pleased, admire & floral, fresh, lemon, chemical & pleased, contented, stimulated \\
Mouthwash & fresh, pungent, mint & stimulated & mint, fresh, medical, pungent & pleased, stimulated, contented \\
\makecell[tl]{Osmanthus \\ Essential Oil} & floral, warm, fresh & pleased & floral, fruity, sweet, fresh & pleased, contented, admire \\
Shoe Polish & floral, fresh, pungent, mint & - & waxy, oily, leathery, earthy & contented, stimulated, admire \\
Laundry Liquid & fruity, grassy, floral, fresh & pleased & floral, fresh, clean & \makecell[tl]{pleased, contented, hopeful, \\ relaxed, calm} \\
\bottomrule
\end{tabular}%
}
\caption{Results of human-chosen versus GPT-chosen odor descriptors and emotion descriptors.}
\label{tab2}
\end{sidewaystable}

\subsubsection{Human olfactory perception}
We aggregated descriptor selections to derive overall patterns (Figure~\ref{fig:user_study_1_results}(a)). It can be inferred that the more frequently a particular odor descriptor was selected, the more common this olfactory experience was among the participants, and further the population. Consequently, we chose the top-three selected descriptors as each odor’s typical and shared olfactory sensation. In cases where descriptors received identical scores, both were retained. Emotion scores were averaged to identify negative (1\textasciitilde3 -- `annoyed,' `melancholic,' `despairing,' `relaxed,' `calm,' and `disgust'), neutral(3\textasciitilde5), or positive ranges(5\textasciitilde7 --`pleased,' `contented,' `hopeful,' `stimulated,' `excited,' and `admire').

Figure~\ref{fig:user_study_1_results}(b) shows scores of each emotion dimension and the second and third columns in Table~\ref{tab2} shows the final results for human-chosen odor and emotion descriptors. The preliminary results provided the descriptive data for odor visualization in our formal study, and demonstrated several mechanisms of human olfactory perception which are briefly summarized as follows:

\begin{enumerate}[(1)]
\item \textbf{Direct sensory stimulation.}
Odors with gentle or fragrant qualities—such as Lemon, Tea, Shampoo, and Shower Gel—commonly evoked positive emotions like `pleased,' `contented,' `hopeful,' and `admire.' By contrast, harsher chemical odors (e.g., Latex Paint, Glue) prompted negative labels including `annoyed,' `despairing,' `stimulated,' or `disgust.' Meanwhile, highly pungent but not inherently dangerous odors, such as Baijiu and Vinegar, were frequently called `stimulated,' suggesting a heightened but not necessarily adverse reaction.

\item \textbf{Odor substances as mediation.}
Participants commonly relied on a substance’s name, key ingredient, or physical state when assigning descriptors. For example, smell of Lemon, Tea, and Coffee were often matched to descriptors including `lemon,' `tea,' or `coffee' respectively. In some cases, the material’s state influenced these choices: P29 described Sesame Oil as `edible,' `liquid,' and `thick,' P25 considered Latex Paint `oily,' and P18 noted Varnish as `paint-like,' `sticky.' When participants failed to identify a substance correctly, they used less-specific terms — for instance, only four recognized `grape,' for Grape, though 21 still categorized it as a fruit. Notably, P2 and P14 conflated Grape with `cantaloupe,' P6 with `papaya,' and P15 with `orange,' reflecting how vision or other cues are often necessary to resolve olfactory ambiguity.

\item \textbf{Association with former life experiences.}
Participants consistently linked certain odors to their prior usage or living contexts, which shaped either negative or positive impressions. For example, Latex Paint, Varnish, and Glue were broadly labeled `pungent' or `chemical,' with some participants (e.g., P1, P5, P12) deeming them harmful or corrosive. In contrast, Tea, Shampoos, Shower Gel, and Detergent — often linked to refreshment or personal care — were perceived as `calm' and `fresh,' reflecting a sense of cleanliness and relaxation (P3, P5, P6, P8, P17, P18, P20, P23, P28).
\end{enumerate}

\subsubsection{LLM’s olfactory perception}
When prompted to select odor descriptors and score emotion pairs, GPT generally provided 3–7 descriptors per odor. Although instructed to choose only from a predefined list, it sometimes suggested synonyms (e.g., `citrus' for `orange,' `acrid' for `pungent') or entirely new terms (e.g., `nutty,' `sweet,' `chemical,' `sharp'). These additional words often aligned with participant feedback — for instance, GPT’s `sweet' for Osmanthus Essential Oil was confirmed by five participants — indicating GPT may have access to a broader conceptual understanding. For emotional descriptors, GPT stayed within the given list and assigned 3–5 per odor, as shown in the fourth and fifth columns in Table~\ref{tab2}. Its overall perception followed three main principles:

\begin{enumerate}[(1)]
\item \textbf{Odor substances as mediation.}
Like human participants, GPT frequently derived descriptors from source names or ingredients. For example, it used `grape' for grape, `tobacco' for Unburnt Cigarette, and `mint' for Mouthwash.

\item \textbf{Associations.}
GPT also considered common olfactory associations that human might have, such as the processing methods and intended uses of the substances. Coffee Bean was tied to `roasted,' reflecting its aroma origins, while Tea was labeled `fresh,' evoking tranquil tea-drinking experiences. Negative associations (e.g., toxic environments) led to descriptors such as `disgust' for glue or latex paint.

\item \textbf{Intensity and pungency.}
Finally, GPT assessed how strong or sharp an odor might be, explaining that highly pungent smells (e.g., Vinegar, Varnish) often yield negative emotional responses. Strong chemical odors such as Mark Pen Ink were similarly deemed `stimulated', `disgust', or `annoyed'.
\end{enumerate}

\subsubsection{Limitations in GPT’s odor perception}
In comparison to human olfactory perception, GPT excels at synthesizing a comprehensive understanding of odors based on its extensive knowledge base. However, lacking real-time sensory capabilities — such as the ability to physically sense odors — GPT’s understanding is inherently more rigid and fixed. Our experiment revealed three key limitations in GPT’s odor perception:

\begin{enumerate}[(1)]
\item \textbf{The uniqueness of odor substances.}
While GPT can provide accurate descriptions for common odors, it encounters difficulties when faced with odors that deviate from the norm. For example, GPT described Unburnt Cigarette as `earthy,' `woody,' and `leathery,' likely reflecting the production process and material properties. In contrast, participants detected a `fruity,' `sour,' and `grape' aroma, with 15 participants specifically mentioning the fruity and sour notes. Participant P1 even likened it to plum candy, which is in sharp contrast to the response of GPT. This highlights GPT’s reliance on generic knowledge, which may not capture more nuanced or idiosyncratic human perceptions.

\item \textbf{Differences in odor concentration.}
GPT’s descriptions tended to emphasize the positive aspects of odors, as seen with Lily Flower, where it selected descriptors like `floral', `fresh', and `sweet'. However, participants often described it as `floral', `pungent', and `earthy', and several mentioned that the intensity of the aroma was unpleasant or overwhelming. Five participants specifically noted that the smell felt foul due to its strong concentration. This discrepancy suggests that GPT’s understanding of odors may not fully account for the variations in perception caused by differences in odor intensity.

\item \textbf{Temporal change of odors.}
GPT also struggles with the temporal dynamics of odors. For instance, in the case of Grape, which were freshly prepared daily but fermented quickly, participants reported detecting hints of fermentation, linking the scent to alcohol. Eight participants specifically mentioned the presence of fermentation or alcohol-like aromas. However, GPT’s descriptions remained focused on `fruity', `juicy' and `fresh', without accounting for the changes in aroma that occurred over time. This indicates that GPT lacks the ability to capture the temporal shifts in odor perception that humans experience.
\end{enumerate}

\section{Formal Study: Participant Evaluation of Odor Visualization}
\label{sec5}

This formal study was designed to evaluate the effectiveness of our odor visualization system. The research was guided by three key goals: (1) to explore the impact of rich and integrated graphical elements on participants' synesthetic experiences (addressing \textbf{RQ2}), (2) to examine the role of language-based descriptors in facilitating odor visualization and triggering synesthetic responses (addressing \textbf{RQ3}), and (3) to investigate the influence of different artistic styles, particularly abstract styles, on the perception of odors and the enhancement of the aesthetic qualities of images (addressing \textbf{RQ4}).

In this study, participants were shown 540 generated odor images representing 20 different odors used in the preliminary study. After viewing each set of images, participants completed a Likert-scale questionnaire and participated in a semi-structured interview. The evaluation focused on two main dimensions: correspondence and aesthetics. Correspondence refers to the extent to which the images conveyed the semantic information of odors, including the substance, olfactory perception, and associated emotions. Aesthetics concerns the overall artistic quality and the potential of the system for not only generating scientifically accurate results, but also for designing visually appealing results.

\subsection{Methods}
\subsubsection{Generating and presenting images}
\begin{figure}[t]
\centering
\includegraphics[width=0.9\columnwidth]{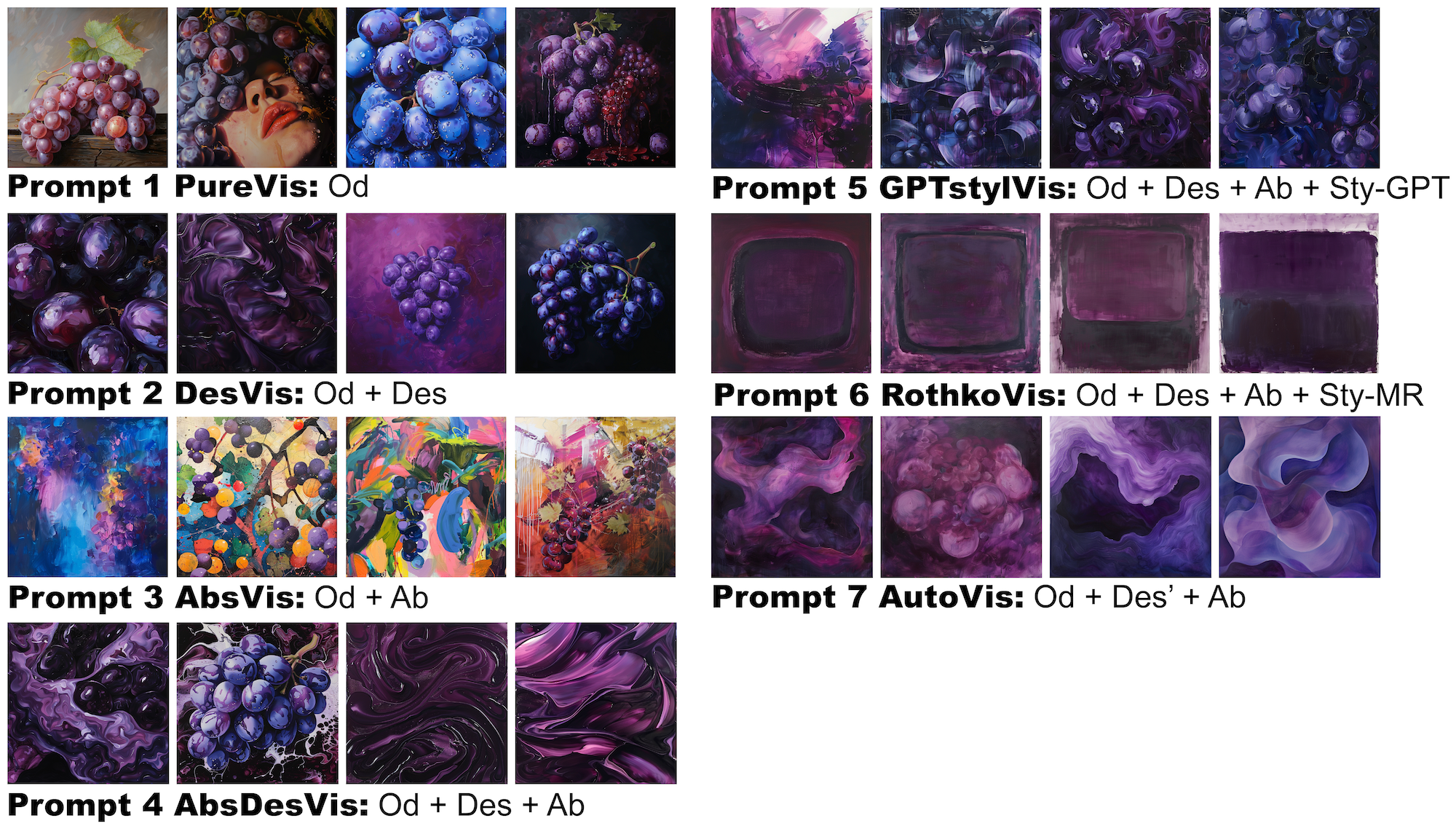}
\caption{Image examples of grape smell. For each prompt, we showed all four variations from one generation. There are obvious differences between each image group. For instance, Prompt 1 \textit{PureVis} illustrates a figurative style, and Prompt 6 \textit{RothkoVis} represents a cubic and color-composed abstract style. Participants were asked to give an overall evaluation based on four images of each style.}\label{fig:grape_examples}
\end{figure}

We used the \textit{Paint by Odor} system to generate seven distinct types of images based on seven different prompts. Each image group was represented by four variations to account for the inherent variability of the Midjourney algorithm. Figure~\ref{fig:grape_examples} shows examples of the odor visualizations for Grape, including the generated images and prompts. The Midjourney version used was v6.0, with all other parameters set to default. In total, each odor was represented by 28 images (7 forms × 4 variations), resulting in 540 images (20 odors × 28 images) for the participants' evaluation. All images are available for review in the \href{https://github.com/gordon7yu/Paint-by-Odor/blob/main/ImageGallery.md}{Image Gallery}.

We presented the images using a custom-designed slide format. Each slide included the following elements: four generated images (centered), the odor substance (above the images, in both Chinese and English), a description of the olfactory and emotional attributes (bottom right of the images, in both languages), an indicator bar to show the form order (seven dots beneath the images), and a progress tracker indicating the study’s current position (left side of the slide).

\subsubsection{Participants}
We recruited 28 participants via social media platforms. To ensure the reliability and validity of the data, we screened participants through a detailed questionnaire that gathered information on factors such as age, gender, educational background, visual impairment, anosmia, interest in art exhibitions, odor sensitivity, and prior experience with generative AI tools. 

We excluded participants with little interest in art exhibitions or poor sensitivity to odors to ensure the participants had the necessary background and sensory sensitivity to engage meaningfully with the odor images. Individuals with low interest in art exhibitions might not fully appreciate or engage with the aesthetic aspects of the system, while those with poor odor sensitivity may not provide accurate or relevant feedback regarding the correspondence between the images and odors. This exclusion criterion helped ensure that the participants' experiences were aligned with the study's goals of exploring both the visual and olfactory aspects of the system.

The final sample consisted of 13 males (46.4\%) and 15 females (53.6\%) with ages ranging from 18 to 54 (mean = 27.3, std = 9.5). All participants self-reported no visual impairments or anosmia. They came from diverse educational backgrounds, including engineering, science, art and design, medicine, management, psychology, and education. All participants received a reward of CNY 120 or an equivalently valued gift. Detailed participant demographics are available in Table~\ref{tabS1} in the Appendix. The experiment was also approved by the Tsinghua University Science and Technology Ethics Committee (Medicine).

\subsubsection{Procedure and questionnaire}
\label{formal_study_questionnaire}
\begin{figure}[t]
\centering
\includegraphics[width=0.6\columnwidth]{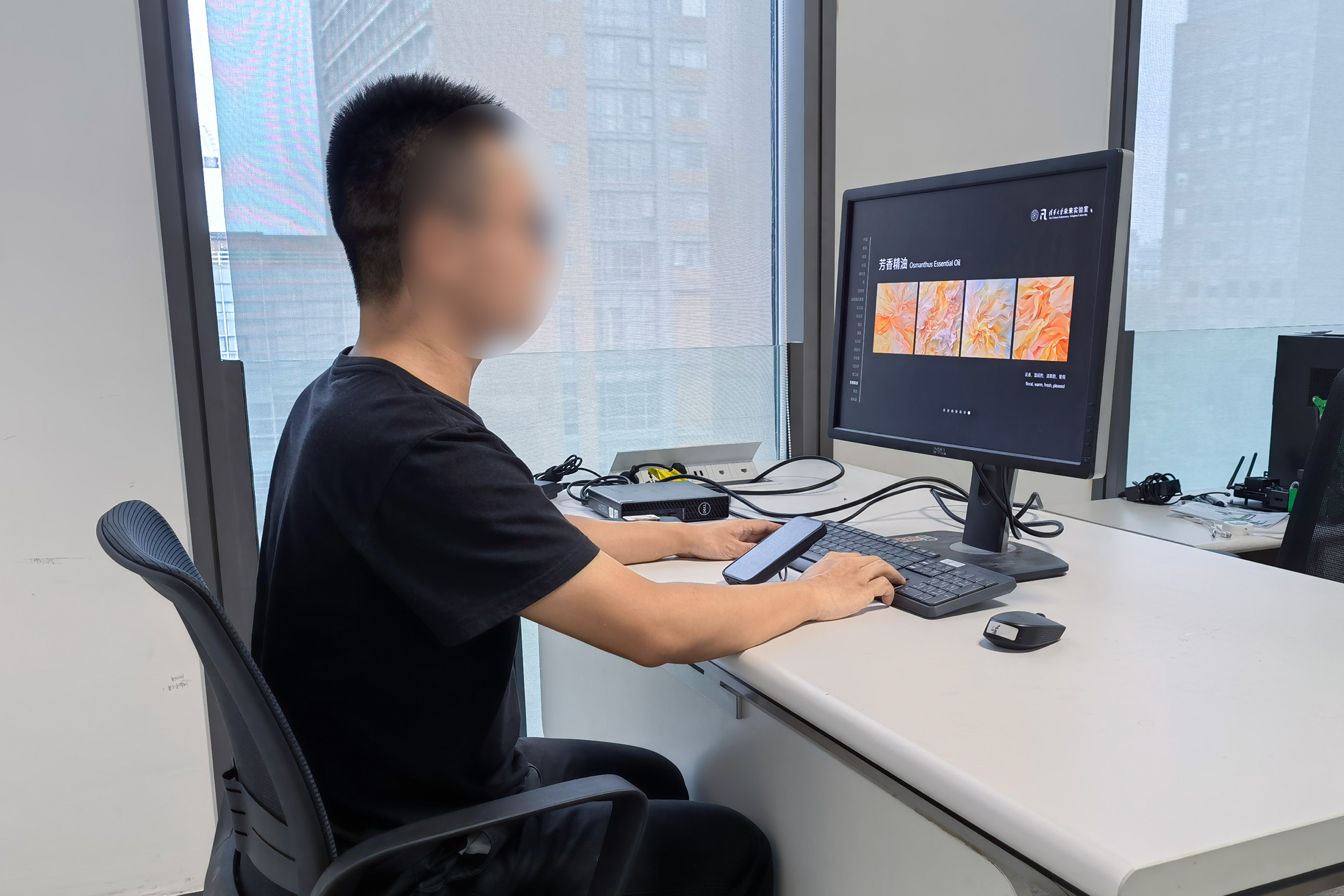}
\caption{Picture of the formal study. Participants were presented with a slide showing all images odor by odor. Within each odor, seven groups of images were presented prompt by prompt. The odor and emotion descriptors from our preliminary study were shown below the images for evaluation.}\label{fig:user_study_2_pic}
\end{figure}

Participants took part in the study either in the lab or remotely. In the lab, participants viewed the odor images on a PC monitor in a controlled 3m x 3m office space. Remote participants were provided with a link to the same image set and viewed the images on their personal computers to avoid quality degradation from screen sharing. The procedure remained consistent for both groups. Figure~\ref{fig:user_study_2_pic} shows the lab setup.

The study lasted approximately two hours per participant. During this time, participants evaluated seven sets of images for each odor, corresponding to seven distinct prompt types, totaling 20 odors. To minimize fatigue, participants were given a 10-minute break after evaluating the first 10 odors.

After viewing each set of images, participants completed a questionnaire with the following questions:

\begin{enumerate}[1)]
\item Do the images reflect the odor source? (Yes/No)
\item How well do the images correspond to the odor description? (5-point Likert scale)
\item How interested are you in these images? (5-point Likert scale)
\item How much do you like these images? (5-point Likert scale)
\item Which emotions do these images evoke? Please choose from the following six pairs of emotion descriptors. (Negative, Neutral, or Positive)
\end{enumerate}

The first two questions evaluated how well the images conveyed odor-related information, addressing our research questions on the impact of graphical elements, language descriptors, and styles on synesthetic experiences. The third and fourth questions assessed the aesthetic appeal of the images, using intuitive indicators of liking and interest to evaluate the artistic quality of the images (\cite{lyssenko2016evaluating}). The fifth question focused on evaluating the emotional synesthesia evoked by the images.

After viewing all seven forms of images for one odor, participants were asked two follow-up questions:

\begin{enumerate}[1)]
\item Which image form do you feel best corresponds to the odor description? 
\item Which image form do you like the most?
\end{enumerate}

These questions allowed participants to reflect on their choices and provide reasoning, offering insight into their evaluation criteria for correspondence and likes. This qualitative feedback was crucial in understanding individual experiences that quantitative data alone could not capture.

At the end of the study, we conducted a final interview to discuss user preferences and potential future applications of the system. The questions included:
\begin{enumerate}[1)]
\item Can you differentiate between the image forms for all 20 odors? 
\item Which image form do you think is most suitable for visualizing odors?
\item What potential applications do you envision for this odor visualization system?
\end{enumerate}

\subsubsection{Analysis}
In this study, we adopted a mixed-methods approach to analyze the collected data, combining both quantitative and qualitative techniques to provide a comprehensive understanding of the research questions.

\textbf{Quantitative Analysis:} The Likert-scale responses were analyzed using descriptive statistics (mean and variance) to assess patterns in participants’ sensory perceptions and aesthetic evaluations across different odors, visual styles, and groups.

\textbf{Qualitative Analysis:} Interview data was transcribed, coded, and categorized based on key dimensions like emotional responses and sensory associations. This helped identify how visual elements influenced participants' olfactory experiences.

\textbf{Mixed-Methods Integration:}
First, we categorized the visual elements mentioned by participants to determine which were most linked to synesthetic experiences. We then re-categorized the data by image sets to evaluate the impact of different styles and prompt units on odor perception and aesthetics. A combination of word frequency analysis and thematic analysis provided deeper insights into the findings.

\section{Results}
\label{sec6}
In this section, we present the results, starting with a summary of the overall quantitative data from questionnaires and interviews. Through coding, we identified key visual elements that contribute to olfactory synesthesia, termed \textbf{odor visualization focus}. We then explored the cross-modal mechanisms underlying each focus. Lastly, we performed thematic analysis to assess the impact of our designed prompt units on olfactory synesthesia and aesthetic experience, concluding with a summary of potential future applications based on participant input.

\subsection{Quantitative results}

\begin{figure}[t]
\centering
\includegraphics[width=0.8\columnwidth]{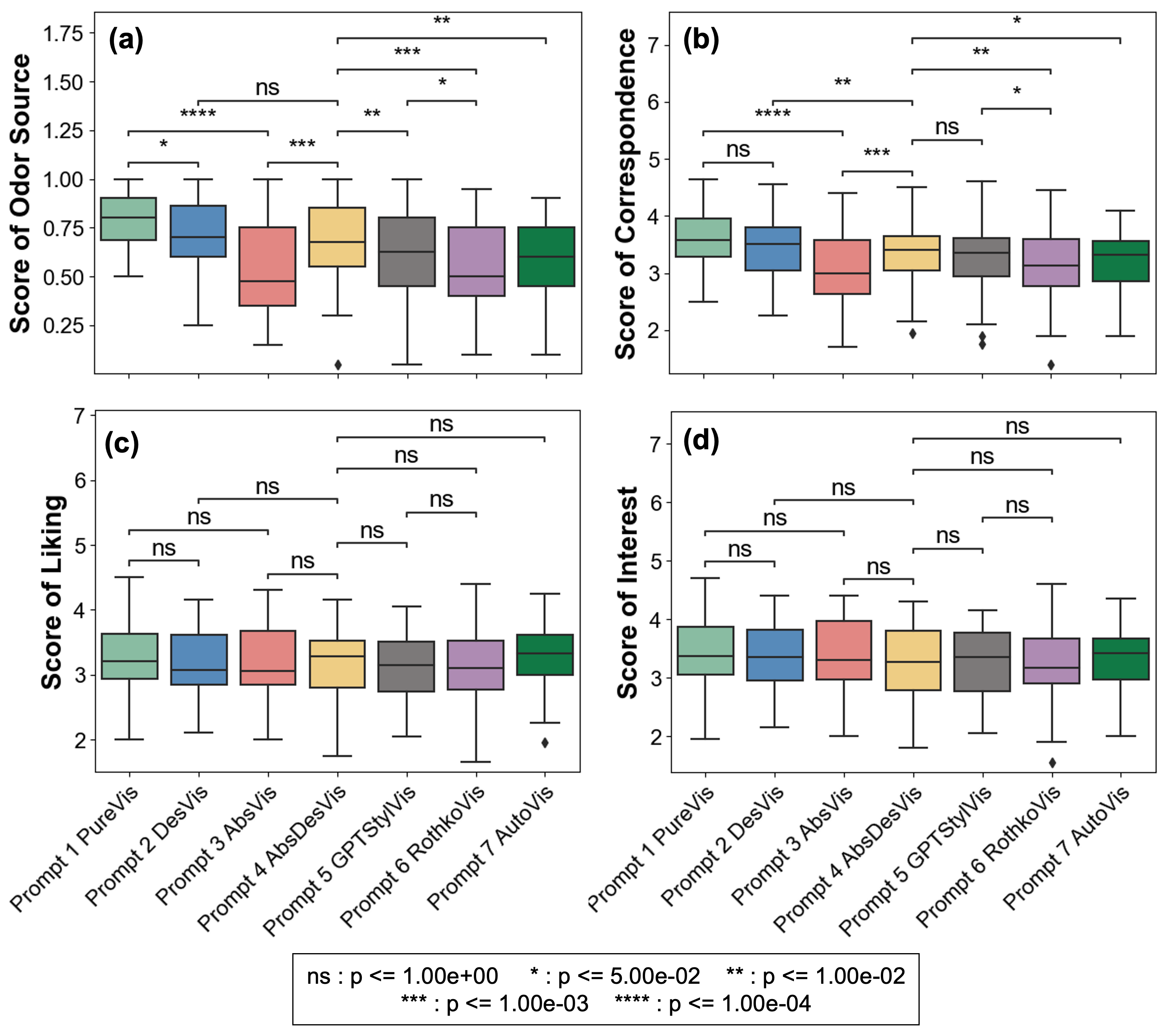}
\caption{Rating results from questionnaires. Significance analysis were labeled only between prompts with one element different. Prompt 1 \textit{PureVis} reflects odor source information the most, and also scored highest on average regarding correspondence. While Prompt 3 \textit{AbsVis} scored the lowest in both dimensions. In terms of liking and interest, the prompts showed minimal variation, with Prompt 7 \textit{AutoVis} slightly outperforming others, while Prompt 5 \textit{GPTstylVis} and Prompt 6 \textit{RothkoVis} performed slightly worse in these dimensions.}
\label{fig:user_study_2_results}
\end{figure}

\begin{figure}[htb]
\centering
\includegraphics[width=1\columnwidth]{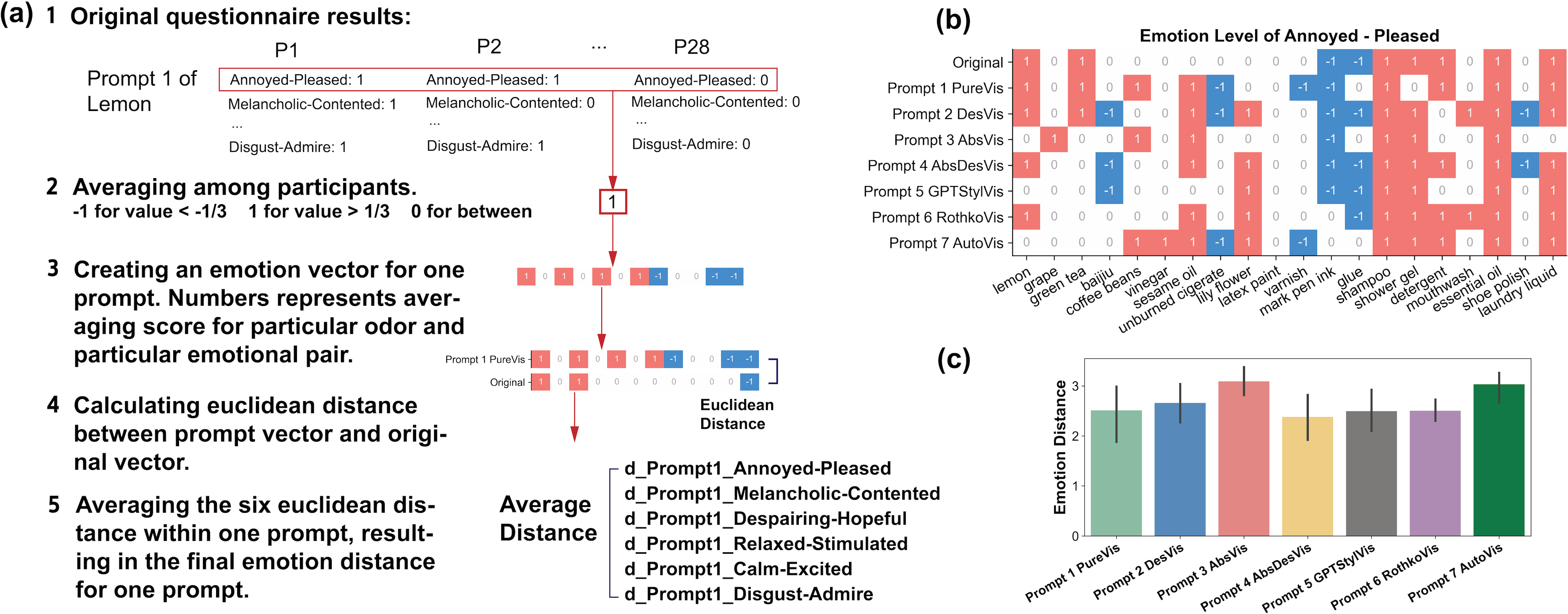}
\caption{Results of emotions. (a) A diagram showing the process of calculating emotion distance of each prompt. (b) An example of emotion average scores of annoyed-pleased pair. (c) Emotion distance results. The lower the score, the closer the generated images are to the original emotions (results of the preliminary experiment).}
\label{fig:emotion_result}
\end{figure}

\begin{figure}[t]
\centering
\includegraphics[width=0.8\columnwidth]{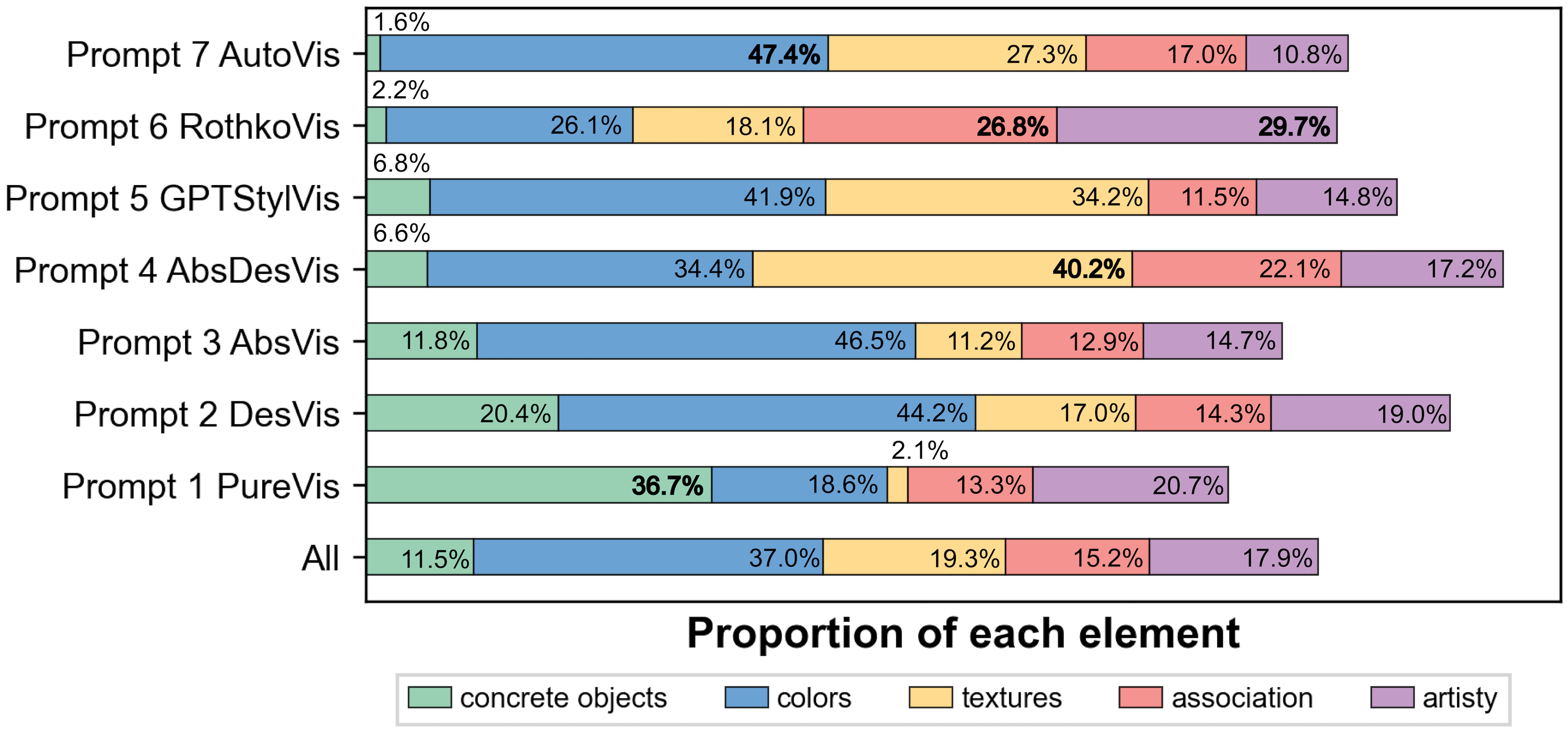}
\caption{The proportion of odor visualization elements within each prompt. ``All" represents the overall distribution proportion of all elements. Concrete elements have the highest proportion in Prompt 1. Under the mediation of \textit{Des'}, colors have the highest proportion in Prompt 7, while textures dominate in Prompt 4. Influenced by Mark Rothko's style, participants generated more associations and experienced greater aesthetic feelings when viewing images of Prompt 6. The bold numbers indicate the highest proportion. This distribution reflects the diverse ways each prompt encodes olfactory perceptions and visual expressions.}
\label{fig:proportion_of_elements}
\end{figure}

\subsubsection{Ratings from the questionnaire}
We first examined the distribution of quantitative data using the Lilliefors test, revealing non-normal distributions for all four questions: odor source evaluation, correspondence score, liking score, and interest score (p \textless 0.05, see Appendix Figure~\ref{fig:norm}). This indicates that user preferences influenced the results. Hence, we applied non-parametric tests, such as the Wilcoxon test, for subsequent analysis.

Figure~\ref{fig:user_study_2_results} presents the average scores. Prompt 1 was most effective at reflecting odor source information and correspondence, while Prompt 3 scored the lowest in both categories. In terms of liking and interest, the prompts showed minimal variation, with Prompt 7 slightly outperforming others, while Prompts 5 and 6 performed slightly worse in these dimensions.

For emotion evaluation, participants rated emotional descriptors on a scale from -1 (negative) to 1 (positive). Figure~\ref{fig:emotion_result} shows that Prompt 4 best reflected the emotional descriptions of odors, while Prompt 3 was the least correlated.

\subsubsection{Odor visualization focus analysis}
Through coding participant interview comments, we identified nine odor visualization focuses, categorized into 5 types: \textbf{concrete objects, colors, image textures (shapes, lines, and textures), associations, and artistry (composition, and art style)}. These focuses represent intermediate links between visual images and olfactory perception and offer valuable design insights. Of the 1076 total codes, 37\% were related to colors, 19.3\% to textures, 17.9\% to artistry, 15.2\% to associations and 11.5\% to concrete elements.

The distribution of these elements varied across different prompts, reflecting the diverse ways in which each prompt encoded olfactory perceptions and visual expressions. Figure~\ref{fig:proportion_of_elements} illustrates the distributions.

For Prompt 1 \textit{PureVis}, concrete elements and related associations were the primary focuses, with 36.7\% of codes related to concrete objects and 13.3\% to associations. 33 of 39 codes(20.7\%) in artistry were expressions of concreteness and realism. 7 participants with 21 codes expressed this style, which emphasized realism, strongly triggered associations with familiar scenes. 

In Prompt 2 \textit{DesVis}, colors and textures became more prominent, with 44.2\% of codes related to colors and 17.0\% to textures. This prompt focused more on color modulation and abstract textures, reducing the emphasis on concrete elements.

Prompt 3 \textit{AbsVis}, which exclusively utilized Gen AI to create abstract paintings for odors, emphasized color as the predominant focus (46.5\%), with concrete elements making up only 11.8\% of the codes. The results indicate that abstraction primarily centered around color creation.

Textures emerged as the dominant focus in Prompt 4 \textit{AbsDesVis}, with 40.2\% of codes referring to textures, followed by colors (34.4\%). Compared to Prompt 2, textures were granted more attention in abstract images than concrete ones. This suggests that the abstract design successfully conveyed olfactory perceptions through texture and color modulation.

Prompt 5 \textit{GPTStylVis}, which used GPT-generated style modulation, showed similar results to Prompt 4, with a slight increase in the emphasis on color (41.9\%) and textures (34.2\%). Some participants noticed artistic styles such as impressionism, but there was no significant consensus.

Prompt 6 \textit{RothkoVis}, the most abstract design, generated more comments related to imagination and artistry (26.8\% and 29.7\%, repectively). This style, inspired by Mark Rothko, was particularly appreciated for its aesthetic qualities, which led to higher liking scores.

Prompt 7 \textit{AutoVis}, a unique style with a distinct color palette, emphasized colors (47.4\%) and textures (27.3\%), creating feelings of cleanliness and relaxation. Comments by participants highlighted the differences in color and texture modulation by GPT (\textit{Des'}) compared to Prompts 2, 4, 6.

Overall, our system successfully encoded olfactory perceptions into images, with each prompt offering distinct visual expressions. The next section will delve into the mechanisms behind these results and their effects on olfactory synesthesia.

\subsection{Mechanisms of Olfactory Synesthesia from Images}
\subsubsection{Individual Visual Elements}

\begin{figure}[h]
\centering
\includegraphics[width=0.75\columnwidth]{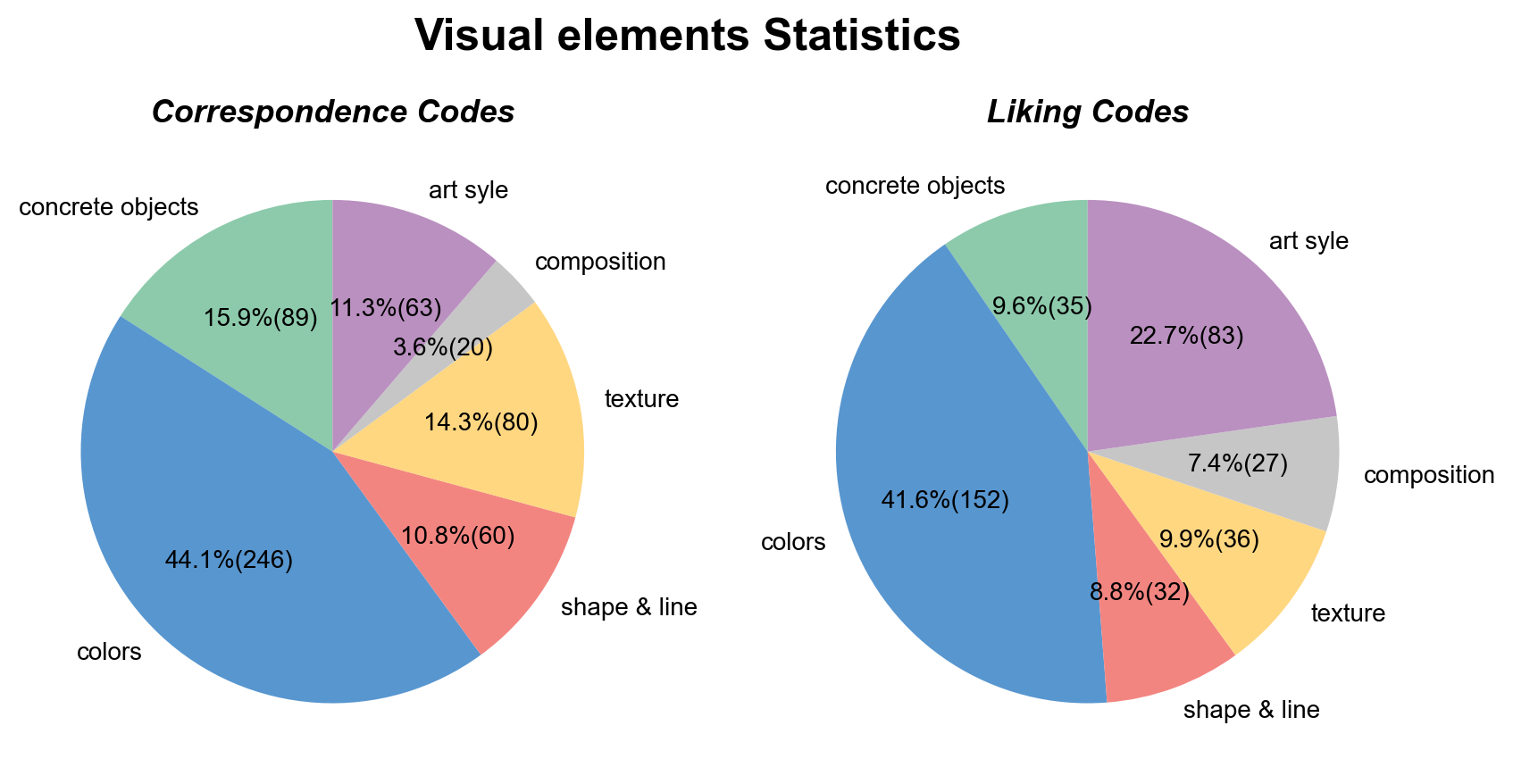}
\caption{Statistics of different visual elements. (a) Correspondence. (b) Liking. Color emerges as the dominant focus in both dimensions, highlighting its key role in conveying olfactory perceptions and enhancing artistry.}
\label{fig:visual_elements_pie}
\end{figure}

\textbf{\emph{Colors}}
\textbf{Color stands out as the most intuitive visual element related to odors}, and the majority of the feedback reflects associations between specific colors and descriptors of odors or emotions (Figure~\ref{fig:visual_elements_pie}). Participants demonstrated consistent patterns in color-odor associations, which are often mediated by emotions and semantics. For example, 26 participants associated green with stimulating emotions, linked to descriptors such as fresh, sour, minty, or pungent. P14 remarked that \textit{``The color green evokes sensations of pungency and fermentation."} Similarly, P8 noted that \textit{``Yellow is associated with fruity scents,"} especially in images depicting detergent odors.

Table~\ref{tab:odor_substances} provides examples of typical odor-color and emotion-color pairs based on the categorized codes. Additionally, the combination of multiple colors enhances the perceived richness and intensity of scents. Seventeen codes point to this idea (mostly from Prompt 3 \textit{AbsVis}), with participants associating rich, vibrant color schemes with strong scent experiences. For instance, P8 commented that \textit{``The variety of colors allowed me to experience fruity, grape, and sour notes in the scent of the cigarette."} Four participants further suggested that rich colors evoke a sense of stimulation, mirroring the intensity of odors. This suggests that \textbf{using colors with high saturation in odor images could enhance the perceived strength of a scent.}

From an aesthetic perspective, color also plays a crucial role in participant preference. Images featuring rich, vibrant colors tended to be favored. For example, P1 stated \textit{``The third group (Prompt 3 AbsVis) is my favorite because its colors are vibrant and striking,"} while P8 expressed \textit{``The seventh set (Prompt 7 AutoVis) of grape images is my favorite because the varying color shades make it more beautiful."} This indicates that the aesthetic appeal of odor visualization images is significantly boosted by the use of rich, dynamic colors.

\begin{table}[htbp]
\resizebox{\textwidth}{!}{
\begin{tabular}{@{}llll@{}}
\toprule
\textbf{Color} & \textbf{Odor Descriptor} & \textbf{Emotion Descriptor} & \textbf{Number of codes} \\
\midrule
orange \& yellow & orange, fruit & pleased & \makecell[tl]{17 (P2, P3, P4, P5, P8,\\P9, P13, P14, P16, P21, P22)} \\
green & \makecell[tl]{fresh, sour, \\ mint, pungent} & stimulated & \makecell[tl]{26 (P1, P3, P4, P5, P7, P8, P9, P11, \\ P12, P13, P14, P20, P22, P25,P26, P28)} \\
red & spicy, soil & stimulated & 10 (P3, P4, P6,P9,P16,P18, P20) \\
brown & roasted, burnt & - & 11 (P1, P3, P6, P9, P14, P18, P20, P22) \\
black & \makecell[tl]{roasted, soil, \\ ink, fresh, pungent} & \makecell[tl]{despairing, annoyed, \\ stimulated} & \makecell[tl]{20 (P1, P2, P5, P6, P7, P11, \\ P13, P16, P18, P20, P24, P25, P27)} \\
blue & fresh, mint & stimulated, despairing & 15 (P4, P7, P11, P13, P16, P18, P25) \\
\bottomrule
\end{tabular}
}
\caption{Participant-reported color-odor pairs.}
\label{tab:odor_substances}
\end{table}

\textbf{\emph{Shapes \& Lines}}
Certain shapes and lines evoke specific synesthetic olfactory responses, indicating strong correspondences with smells. Although less frequent than color associations, shapes and lines still play a significant role in odor visualization. These include:

\begin{enumerate}[1)] 
\item Curved lines are linked to juicy and fermented sensations, as seen in Grape and Sesame Oil images. For example, P3 expressed the following regarding the grape image: \textit{``The fluidity of the image gave me the impression of something juicy.''} 
\item Sharp lines corresponded to pungent, spicy, and stimulating odors, as in Vinegar, Mar Pan Ink, Lily Flower, and Mouthwash. P11 commented, \textit{``Sharp edges are pungent and spicy.''} \end{enumerate}

For example, \textbf{Participants tended to associate streamlined and curved forms with smoothness, freshness, fluidity, and excitement.} Thirteen codes reflect this trend. P26, when discussing Shampoo images (Prompt 4 \textit{AbsDesVis}), said, \textit{``I like the smooth sensation conveyed by such imagery.''} Conversely, square shapes often evoke feelings of limitation and control, which some participants found unappealing. P6, for instance, disliked the grape images in the \textit{RothkoVis} group, stating, \textit{``They cannot make me associate them with fermentation. Squares represent limitation and control.''} Thus, certain shapes \& lines have their own special representations of either odor form or emotions and feelings. Appropriately using these elements can enhance both aesthetic value and synesthetic coherence of odor visualizations.
 
\textbf{\emph{Textures}}
\textbf{The connection between textures and odor perception stems from the material qualities of the odor source.} Participants often associate the texture of a visual element with the tactile qualities of the odor substance itself. For example, images of Latex Paint were frequently noted for their sticky texture, reminiscent of the odor source. P3 commented, \textit{``The texture gives a feeling of oil.''} This suggests that using textures that mirror the physical characteristics of the odor source can strengthen the image-odor relationship.

\textbf{Brushstroke techniques, although contributing to aesthetic appeal, are highly subjective.} Different participants had varied preferences for painting styles, such as oil painting or blending. P18, when discussing the Lemon odor images (\textit{AbsVis}), remarked, \textit{``It looks very beautiful and artistic, with a sense of oil painting,''} while P10 appreciated the blending effect in the Green Tea images (\textit{AutoVis}), stating, \textit{``I really like the feeling of the blending.''} P22 noted that the varnish images (\textit{RothkoVis}) conveyed a sense of concealment, with colors \textit{“faintly revealed.”}

\textbf{\emph{Concrete Objects}}
Concrete objects refer to realistic depictions of odor-related substances, scenes, or objects such as liquid drops and fumes. For instance, images of half-cut lemons and green leaves often evoke a fresh sensation. Fumes, in particular, are a powerful visual representation of odors. P9, commenting on Varnish images (\textit{PureVis}), said, \textit{``Smoke and fog are linked to the nauseating smell of burnt oil.''} Similarly, P28, discussing Baijiu images (\textit{AutoVis}), said, \textit{``The bubbles and fermentation excitement are similar to the visual sensation of looking at Baijiu in a glass.''} Using concrete elements tied to the odor source can enhance the connection between the image and the olfactory experience.

\textbf{\emph{Composition}}
Participants indicated that well-composed images were more aesthetically pleasing. However, there was no consensus on what constitutes good composition, suggesting individual differences in aesthetic preferences. For example, P11 preferred \textit{GPTstylVis} Vinegar images for their \textit{``graphical richness,''} while P13 favored \textit{AbsVis} Essential Oil images for their \textit{``traditional Chinese painting techniques''} with more blank space. In terms of composition, balancing contrast, richness, arrangement, and blank space can enhance the visual appeal and clarity of odor-related images.

\textbf{\emph{Art Styles}}
In our study, the participants could identify various artistic styles in the images, including abstraction, minimalism, modernism, and others such as children's drawing, oil painting, and decorationism. The experience of art styles is an expression of artistic appreciation. Incorporating distinct art styles into odor visualizations can elevate their artistic value. 24 participants provided feedback on art styles, with 146 related codes. Like composition, these comments primarily appeared in response to preference questions.

\subsubsection{Imagination and Association}

This study revealed that imaginative and associative processes play a vital role in eliciting olfactory sensations from visual images. Through coding analysis, we identified three categories of associations:

\textbf{Object Association.}
Object association refers to participants’ olfactory synesthesia triggered by visual elements that are not explicitly depicted but rather imply specific objects. We collected 70 related codes. For instance, four participants (P3, P5, P7, P27) reported that images of Latex Paint reminded them of walls, while three participants (P5, P17, P24) associated the images of Varnish with wood.

\textbf{Scene Imagination.}
Scene imagination involves constructing stories or scenarios around an odor. Eighty-eight codes reflected this tendency. For example, P8 described the images of Baijiu as follows: \textit{``Each picture features a glass of wine with its open lid evoking the sensation of the stimulating aroma of the wine and the atmosphere of people drinking together.''} The latter description was a story-like scene that usually happens when friends or families gather together. Similarly, P22 remarked, \textit{``There should be spots on the cloth after rubbing shoe polish''} when viewing images of Shoe Polish (\textit{AbsVis}). Such scene-based associations often serve as mediating factors for olfactory synesthesia, particularly when linked to personal memories.

\textbf{Cultural Association.}
Certain odors, such as Baijiu, Green Tea, and Mark Pen Ink, carried cultural significance. Participants (e.g., P25) drew on Chinese cultural references, associating Baijiu with Li Bai’s\footnote{A renowned Chinese poet celebrated for his bold and unrestrained style and his love of drinking.} poetic style. These culturally rooted associations deepen olfactory synesthesia, emphasizing the role of cultural background in shaping odor perception.

\subsection{Analysis of prompt elements}
Our first design principle integrates color, shapes, and lines from an expert GPT model to provoke olfactory synesthesia. This section examines how participants noticed these elements and how effectively they conveyed odors.

\subsubsection{Figurative or Abstract? (Ab, Sty-MR, Sty-GPT)}
Our results show that figurative-style images (Prompts 1 and 2, without \textit{Abs} elements) achieved significantly higher correspondence scores than abstract-style images (Prompts 3–7, \textit{Abs} elements added; Wilcoxon Rank Test, statistic = 5101, p \textless 0.005). However, there were no significant differences in liking or interest (p = 0.542 and p = 0.388, respectively), suggesting that both styles elicited comparable levels of artistic engagement. Despite their lower correspondence scores, abstract prompts attracted more discussion overall; we recorded 345 codes for five abstract prompts (69 codes/prompt) versus 83 codes for two figurative prompts (41.5 codes/prompt).

Several participants highlighted the distinct appeal of abstraction. For instance, P13 described \textit{AbsDesVis} images of Mark Pen Ink as \textit{``a sense of contemporary art, of order within the chaos,''} while P28 favored Prompt 6 of Lemon for its \textit{``simple and minimalist style.''} P15 likened \textit{RothkoVis} images of Shampoo to \textit{``fashion in the streets of Paris''} underscoring the imaginative responses that abstract styles can inspire. P11 observed, \textit{``There’s a rhythm inside the picture, a feeling of hope and vitality.''} (\textit{AutoVis}).

\begin{figure}[t]
\centering
\includegraphics[width=1\columnwidth]{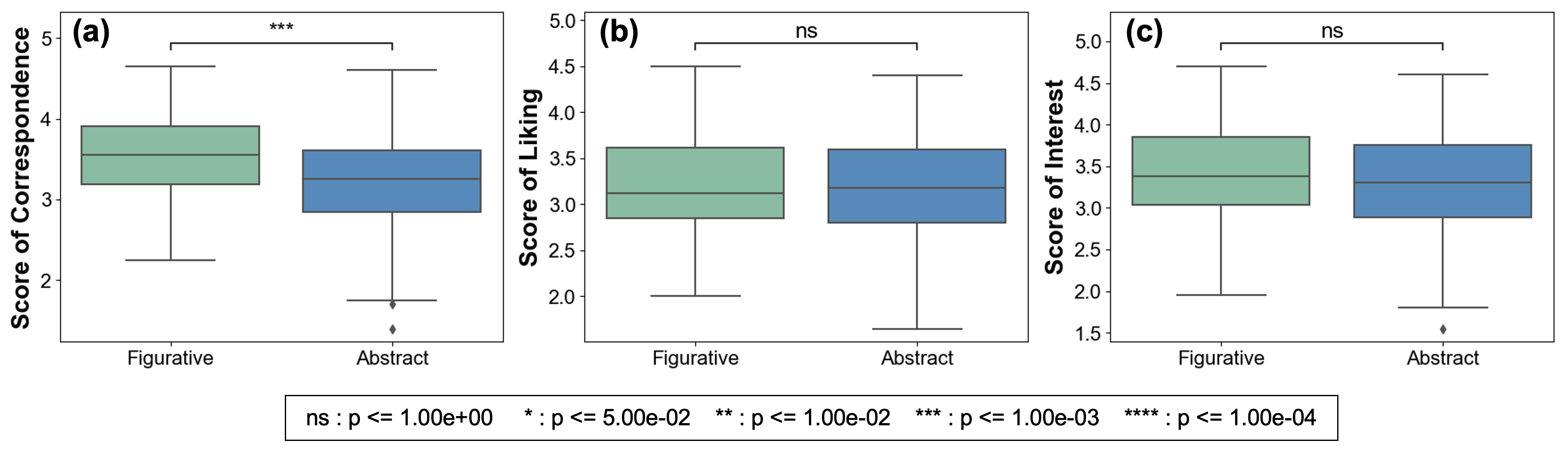}
\caption{Comparison of figurative and abstract prompts for (a) correspondence, (b) liking and (c) interest.}
\label{fig:figurative_vs_abstract}
\end{figure}

\textbf{Influence of Educational Background}
Further analysis showed that participants with art and design related backgrounds rated abstract styles higher than figurative ones (Table~\ref{tab:Pbackgrounds}). Specifically, 55.6\% (10/18) of those who preferred abstract prompts had artistic training, whereas only 18.2\% (2/11) of those who favored figurative styles came from such a background. Individual comments also reflected these differences: P2 (Computer Science) noted, \textit{``I dislike abstract paintings because they do not have clear logic and themes,''} whereas P28 (Modern Art) remarked, \textit{``I studied modern art... so I have a good impression of the modern abstract style.''}

\begin{table}[h]
\begin{tabular}{@{}ll@{}}
    \toprule
    \textbf{Figurative Prompts} & \makecell[tl]{P2 \textcolor{cyan}{\textbf{P3}} P4 P6 P8 P10 P15 P21 P25 \textcolor{cyan}{\textbf{P26}} P11} \\
    \midrule
    \textbf{Abstract Prompts} & \makecell[tl]{\textcolor{cyan}{\textbf{P1}} P5 \textcolor{cyan}{\textbf{P7}} P9 P11 \textcolor{cyan}{\textbf{P12 P13 P14}} P16 P17 P18 \\ \textcolor{cyan}{\textbf{P19 P20}} P22 P23 \textcolor{cyan}{\textbf{P24 P27 P28}}} \\
    \bottomrule
\end{tabular}
\caption{Favorite prompt of each participant. Bold cyan represents participants with art\&design related backgrounds.}
\label{tab:Pbackgrounds}
\end{table}

\textbf{Spotlight on Prompt \textit{RothkoVis}}
Among the abstract prompts, Prompt \textit{RothkoVis} (emphasizing the style of a certain artist) stood out for its \textit{``square style''} (P16) and \textit{``cubic colors''} (P28). It elicited diverse imaginative content — P13 saw \textit{``the reflection of sunrise,''} P19 pictured \textit{``walls of a street café,''} and P26 likened Detergent to \textit{``a kind of cocktail.''} However, some participants found this style less suitable for certain odors that require more fluid representation (P10, P11). Even so, \textit{RothkoVis} performed exceptionally well for specific odorants (shown in Figure~\ref{fig:Prompt6_results}), including Unburned Cigarette, Latex Paint, and Varnish, likely because its dry, geometric aesthetic complemented these substances’ intrinsic characteristics (P2, P3, P17, P19, P23).
\begin{figure}[t]
\centering
\includegraphics[width=1\columnwidth]{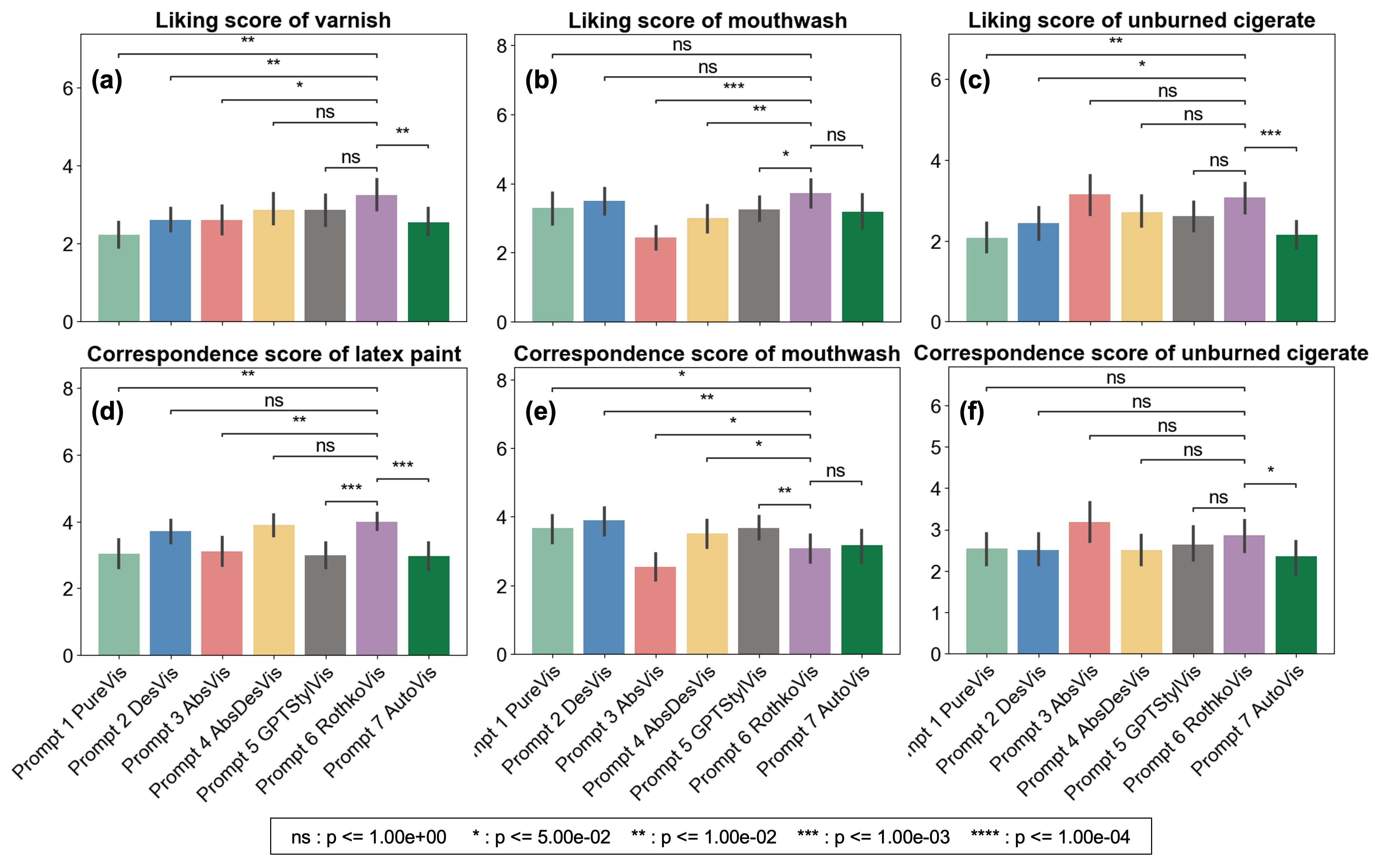}
\caption{Quantitative results showing Prompt 6’s higher correspondence/liking for selected odors. Significance was labeled between Prompt 6 and other prompts.}
\label{fig:Prompt6_results}
\end{figure}

\begin{figure}[t]
\centering
\includegraphics[width=1\columnwidth]{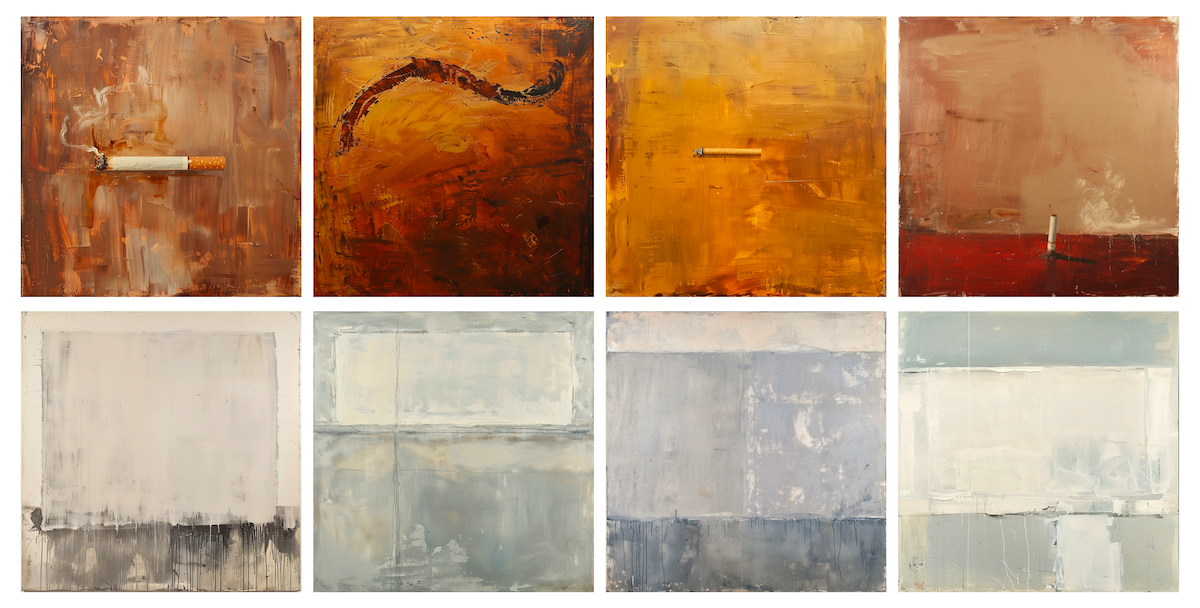}
\caption{Images of Prompt 6 \textit{RothkoVis} for Unburned Cigarette (upper line) and Latex Paint (lower line). The images express unique aesthetic characteristics (from participants' reviews), especially feelings of calmness, peacefulness, and imagination.}
\label{fig:Prompt6_results}
\end{figure}

\textbf{Weak Effect of StyGPT}
According to participants’ subjective reports on distinguishing between prompts (Question 8 in Section~\ref{formal_study_questionnaire}), six participants (P6, P7, P10, P15, P16, P19) explicitly stated that they could not differentiate between Prompt 4 (\textit{AbsDesVis}) and Prompt 5 (\textit{GPTStyVis}). Consistent with this, the scores for correspondence (p = 0.054), liking (p = 0.133), and interest (p = 0.375) revealed no significant differences between these two prompts (Wilcoxon Signed Rank Test). This suggests that the style generated by Expert GPT, names of art schools, did not significantly influence the images within the context of odor visualization. In contrast, compared to the strong effects of \textit{RothkoVis}, it can be inferred that incorporating artist names may have a more substantial effect on the style of images generated by Midjourney, although potential copyright concerns may arise.

In summary, participants found both figurative and abstract styles aesthetically engaging. Figurative images generally provided clearer odor correspondences, but abstract visuals provoked richer discussions and imaginative interpretations — particularly among those with artistic training. These findings suggest that participant background and context-specific factors should guide the choice between figurative and abstract approaches in odor visualization.

\subsubsection{The Effect of Odor Description (Des)}

We tested two prompt pairs that differed only by the inclusion of odor descriptions (\textit{Des}). The first pair was concrete (Prompt 1, \textit{PureVis}, vs. Prompt 2, \textit{DesVis}), and the second pair was abstract (Prompt 3, \textit{AbsVis}, vs. Prompt 4, \textit{AbsDesVis}).

In terms of correspondence, \textit{PureVis} received a higher score than \textit{DesVis} (Wilcoxon Signed Rank Test, statistic = 77, p \textless 0.05), while \textit{AbsDesVis} scored higher than \textit{AbsVis} (Wilcoxon Signed Rank Test, statistic = 19, p \textless 0.0005), as shown in Figure~\ref{fig:user_study_2_results}. These results suggest that the \textit{Des} element is critical for olfactory synesthesia, particularly in abstract contexts. The results for emotional distances further supported these findings, as shown in Figure~\ref{fig:emotion_result}(c). In terms of emotional resonance, Prompt 1 demonstrated a smaller emotional distance from the original stimuli compared to Prompt 2, while Prompt 4 showed significantly closer emotional correspondence than Prompt 3. This indicates that both olfactory and emotional data underscore the greater importance of Prompt Unit \textit{Des} in conveying olfactory information in abstract styles compared to concrete styles. In conclusion, the effectiveness of \textit{Des} further demonstrates the validity of using our Expert GPT system for odor visualization.

\textbf{Qualitative Findings and Potential Explainations.}
An analysis of qualitative data revealed several underlying factors. In the comparison between \textit{PureVis} and \textit{DesVis}, \textit{PureVis} generated images emphasizing concrete, odor-related objects, making it easier for participants to link the images to actual smells (P8, P24). By contrast, \textit{DesVis} added odor and emotional descriptions on top of the concrete prompts, leading participants to focus more on colors and textures (P6, P9). While this broadened the representation of the odor, it also reduced the prominence of tangible objects in the prompt, thereby slightly weakening the direct connection to the smell (P3, P16, P26).

In the case of \textit{AbsVis} and \textit{AbsDesVis}, introducing abstraction removed concrete objects entirely, so olfactory experiences depended on other visual elements such as colors, shapes \& lines, and textures. Here, \textit{Des} served as a more precise source of odor information that effectively guided image generation, resulting in clearer odor associations (P3, P8, P11, P12, P16, P19). As P16 remarked, \textit{``(AbsVis images) have too much color and are not precise enough for conveying smells and emotions… (AbsDesVis images) have more thematic features.''}

\subsubsection{AI vs. Human: Can AI Visualize Smells Independently?}
Prompt 7 (\textit{AutoVis}) was designed to generate images purely from GPT’s olfactory perceptions (\textit{Des'}), basic odor source information (\textit{Od}), and an abstract style (\textit{Abs}), with no human involvement. This approach aimed to assess the feasibility of fully automated odor visualization.

A total of 10 participants described the image set of \textit{AutoVis} (Group 7) as a \textit{``unique style among all groups''} (P1, P2, P6, P9, P12, P13, P15, P16, P20, P26). For instance, P2 mentioned, \textit{``Group 7 is cooler and more concise,''} while P5 noted, \textit{``The color of group 7 is different from the rest.''}

Comparison of Prompt 4 (\textit{AbsDesVis}, which included human-generated descriptions) and Prompt 7 revealed slightly lower correspondence for Prompt 7 (Wilcoxon, statistic = 94.5, p \textless 0.05). However, liking and interest were similar (p \textgreater 0.05), indicating comparable aesthetic engagement
\begin{figure}[t]
\centering
\includegraphics[width=0.6\columnwidth]{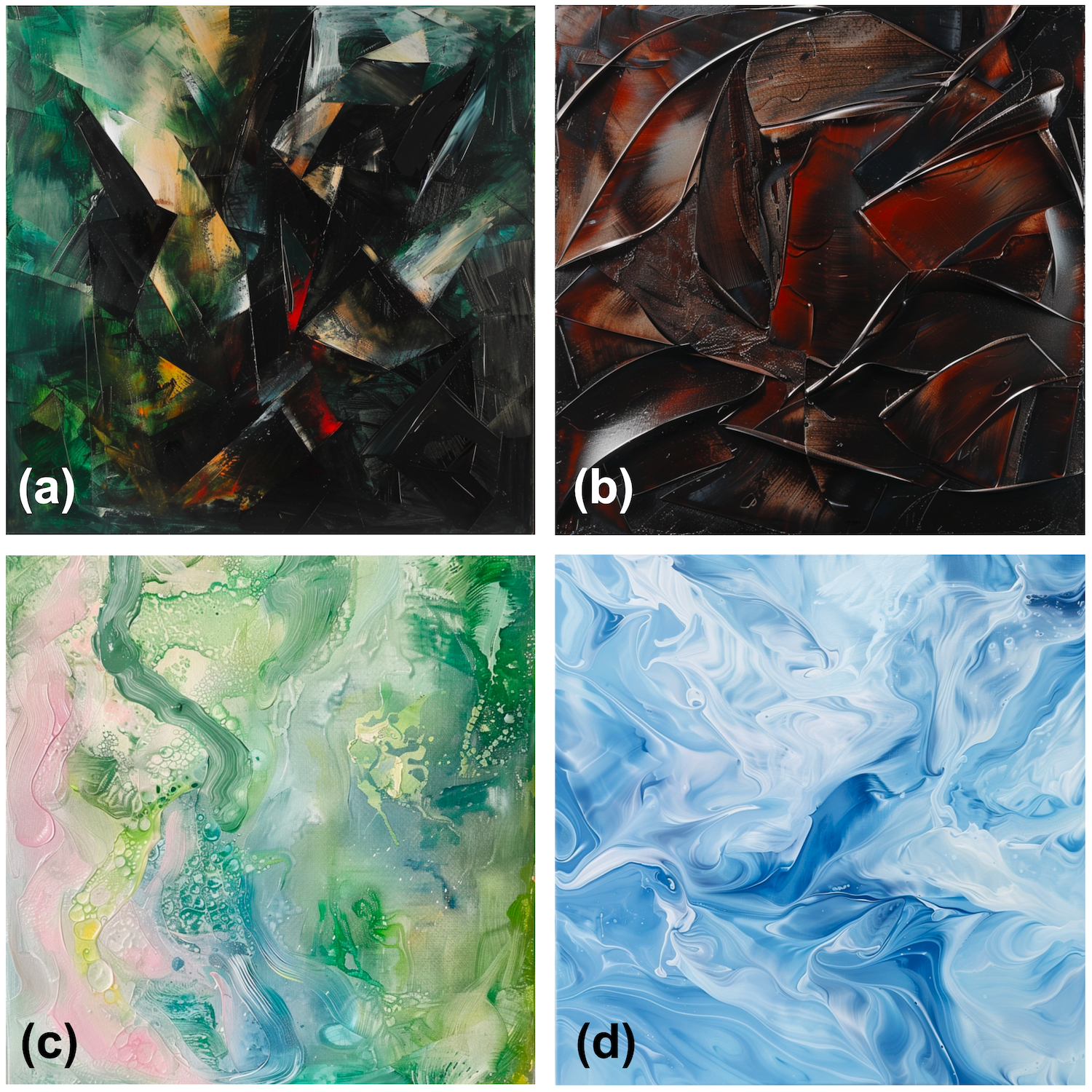}
\caption{Comparison of human-description based images and LLM-description based images. (a) \textit{AbsDesVis} vs. (b) \textit{AutoVis} generated images of Shoe Polish odor, and (c) \textit{AbsDesVis} vs. (d) \textit{AutoVis} generated images of Laundry Liquid odor.}
\label{fig:images_human_vs_gpt}
\end{figure}

Several participants remarked on the rigid or stereotypical representations in \textit{AutoVis}. P3 found it \textit{``somewhat off-topic, presenting stereotypes of smell perception,''} while P17 pointed out that \textit{Shoe Polish} images aligned only with \textit{``brown and black shoe polish in my memories.''} P24 concurred, stating \textit{``My impression of shoe polish is always black and pungent.''} By contrast, Laundry Liquid often appeared in \textit{``blue tones''} (P17), aligning with GPT’s generalized odor knowledge but occasionally conflicting with participants’ personal experiences, as indicated in our preliminary study.

Despite these limitations, fully automated approaches remain promising for odor visualization. Although AI-based prompts may miss the nuanced complexity of human descriptions, their ability to function independently could support scalable, on-demand odor-image generation, particularly when real-time human input is unavailable.

\subsubsection{Potential Applications}
All 28 participants offered various perspectives on future uses for our odor visualization system. Their suggestions span ten key themes:

\begin{itemize}
    \item \textbf{Commercial Design and Sales (22 participants).} 
    Many saw potential for AI-generated images to convey both odors and associated emotions, thereby enhancing consumer engagement. Suggested products included perfumes, shampoos, food, tiles, paints, and mobile phone cases. For example, P2 noted, \textit{``Compared to listing ingredients, an image representing the scent of perfumes is much easier for consumers to understand,''} while P22 suggested it could \textit{``stimulate the desire to buy.''}

    \item \textbf{Spatial Experience Design (8 participants).} Participants envisioned using the system in hotels, restaurants, stations, and vehicles to create multisensory environments by integrating odors with audio or lighting effects. P8 suggested, \textit{``In public stations, the visualization of smells could alert passengers to smoking or other unpleasant odors.''}

    \item \textbf{Assisting People with Disabilities (7 participants).} This theme focused on aiding individuals who have lost their sense of smell. The system could convey important odor-related information, such as the \textit{``beauty of odors and hazardous environmental gases''} (P26), as well as assist in \textit{``training and recovering from anosmia''} (P12).

    \item \textbf{Exhibition Design (5 participants).} For exhibition designers, the system could allow the creation of more immersive experiences by combining odors with visual elements. P25 commented, \textit{``It would be wonderful to both smell and see in a museum setting.''}

    \item \textbf{Dining Experience (4 participants).}  
Suggestions included using the system in restaurants to visualize food flavors, thus heightening the dining experience. P7 envisioned \textit{``a picture of my dinner generated by odor descriptions,''} while P22 said it \textit{``could stimulate diners’ appetites.''}

    \item \textbf{Children’s Learning (2 participants).}  
Participants saw opportunities to teach children about smells, developing their sensory awareness and language skills. P10 proposed training \textit{``children’s cognition of smells and colors, especially for those with cognitive disorders.''}

    \item \textbf{Research Tools (2 participants).}  
Participants suggested using the system in cross-modal smell perception research and fundamental olfactory studies. P15 remarked, \textit{``We could bypass language descriptions and use imagery elements to understand smells,''} while P13 saw potential in \textit{``contributing to fundamental theories by describing smells with colors.''}

\item \textbf{Literary Creation (1 participant).}  
P2 suggested that the system could be used to \textit{``inspire novel writers by generating images based on characters’ personalities and body odors.''}

\item \textbf{Virtual Reality (1 participant).}  
P11 suggested modeling VR scenes by fusing olfactory and visual experiences.

\item \textbf{Social Experiments (1 participant).}  
P14 highlighted \textit{``the charm of translating smells into images''} to spark \textit{``imagination and interpretation,''} potentially employing the system in social experiments.
\end{itemize}

\section{Discussion}
\label{sec7}
\subsection{Future directions for applications}
This study focuses on the mechanistic exploration of odor visualization, presenting the \textit{Paint by Odor} framework primarily as a research tool rather than a finished interactive system. Depending on the intended application, future developments could branch in two directions:

\textbf{Tool-Oriented Systems}
These may include research platforms for investigating odor-visual synesthesia, design tools for commercial products and personalized experiences, and educational or therapeutic solutions for children or individuals with specific olfactory needs. In such contexts, iterative refinement and user personalization becomes vital. For example, in personalized perfume design, prompts might be tailored to users’ olfactory perceptions or scent-related memories, enabling experiences that resonate more deeply with individual preferences. If the system is meant for broader audiences, generalized odor-visual associations may suffice, whereas personalized scenarios can prioritize user-specific inputs and memory triggers.

\textbf{Display-Oriented Systems}
Focusing on physical installations, display-oriented applications aim to convey olfactory information and enhance spatial aesthetics. Environments such as galleries, retail spaces, or smart homes could integrate \textit{Paint by Odor} to enrich multisensory experiences, reinforcing the immersive qualities of the space.

\subsection{Improvement for Paint by Odor}

\textbf{Integrating Users’ Subjective Modulation} In this study, we employed various prompt units to generate different odor-visualization methods, uncovering core mechanisms of olfactory-to-visual transformations and deriving design guidelines. Nevertheless, user diversity and subjective factors remain crucial. Individuals interpret odors through personal experiences and memories, which can substantially influence the final visualization. Future interactive systems might incorporate users’ personal content — such as emotional associations or narrative contexts — to produce more meaningful and customized odor images.

\textbf{Incorporating Electronic Nose Sensing Modules} Although LLMs and Gen AI show promise for fully automated odor description and visualization (e.g., Prompt 7), they currently lack real-time olfactory sensing. Moreover, identifying odor sources (like Lemon) still relies on user input rather than automated detection. To address these gaps, next-generation systems could integrate electronic nose (e-nose) technologies, typically combining gas sensor arrays to detect and output real-time information on odor sources and intensity. By coupling LLMs with genuine sensor data, an AI-driven platform might be able to analyze and visualize complex, evolving odors, closely mirroring human olfactory perception and adapting to changing environments.

\subsection{Process of artistic understanding}

Our findings highlight not only the diversity of olfactory synesthesia but also differences in style preferences and aesthetic engagement. Building on Michael Parsons’ five-stage theory of aesthetic development~(\cite{parsons_how_1990}), it appears that participants vary in how they interpret and evaluate art. Early-stage assessments often involve personal associations with concrete objects or familiar scenes (e.g., \textit{home furniture} and \textit{drinking with friends}), whereas more advanced stages hinge on formal properties such as color, texture, and composition.

This spectrum of aesthetic judgment was evident when participants explained their preferences for certain prompts. Some emphasized subjective, memory-driven connections, aligning with Parsons’ earliest stage, whereas others discussed structural or compositional aspects, reflecting later-stage cognitive processes. Consequently, aesthetic experiences with odor-based images can differ widely, shaped by individual sensory, emotional, and conceptual frames of reference. These variations underscore the complexity of both olfactory perception and art appreciation, influenced not only by personal experiences and emotional states but also by distinct levels of aesthetic understanding.
\section{Limitations}
\label{sec8}
\subsection{Image Generation and Consistency of Gen AI}

Although AI platforms like Midjourney have significantly advanced the controllability and creativity of image generation, notable limitations persist. One primary challenge involves maintaining logical coherence in figurative images. For example, participants frequently mentioned the misrepresentation of Unburned Cigarettes: some images illustrated them burning or lit from the wrong end, which broke immersion and led participants (e.g., P19) to rate those images lower. Additionally, semi-transparent or transparent substances (e.g., Glue and Mouthwash) are difficult to represent accurately through color or texture, causing the output to fail in conveying their inherent translucence or transparency. Future work on modeling light, reflection, and other visual cues may help capture these elusive qualities.

Another constraint lies in the consistency of both LLM-based olfactory perception and artistic styles within a single prompt form. Our study asked the LLM for odor descriptors twice, yielding discrepancies (see Table~\ref{tabS3} in Appendix). Moreover, several participants noted style inconsistencies across the four generated images of a single form. P3 described group 7 as \textit{``not consistent,''} P6 felt each glue image \textit{``had a different style,''} and P19 observed that \textit{``Prompt 3 of Green Tea varied in styles, like random generation.''} These mismatches sometimes led participants to rate an entire form based on one especially appealing or unappealing image. Hence, while generative AI can create visually striking individual outputs, ensuring stylistic stability across multiple images remains an ongoing challenge.

\subsection{Cultural Variations}

Cultural and geographical backgrounds heavily influence olfactory perception and visual-olfactory correspondences. In our experiment, the portrayal of Vinegar illustrated this effect. Although Chinese cuisine often uses Vinegar ranging from colorless or light yellow (Rice Vinegar) to dark brown (Chen Cu), AI-generated images for Vinegar leaned toward red and green tones—likely referencing Western variants such as red wine vinegar or tarragon-infused vinegar. This discrepancy underscores how blended cultural references in LLM outputs can reshape or misinterpret odor imagery.

Because this study exclusively involved Chinese participants, our findings and visual-olfactory mappings reflect a distinct cultural context. Future research should examine similar odor-image correspondences in different cultural settings, each with its own odor vocabularies, culinary traditions, and historical practices, to develop more universally adaptable and culturally nuanced odor visualization systems.

\section{Conclusion}
\label{sec9}
In this work, we introduced \textit{Paint by Odor}, a framework designed to transform olfactory perceptions into visual representations through generative AI (Gen AI) and large language models (LLMs). By systematically composing six prompt elements — such as odor and imagery descriptions and abstraction styles — into seven prompts, we examined how specific visual cues could evoke and enhance users’ olfactory synesthesia. Our preliminary experiment produced olfactory descriptive data and investigated the capabilities of LLMs in producing olfactory perception. The findings indicate that GPT can demonstrate a comprehensive but more rigid perception of odors. The subsequent formal experiment revealed the interplay between odor visualization, aesthetics, and participant experiences. The results show that figurative images achieved superior olfactory correspondence, while abstract visualizations attracted more discussion. Language-based descriptions proved instrumental in translating olfactory descriptions into visual elements, and style preferences were shaped by personal factors, reflecting the inherently subjective nature of smell. Furthermore, the fully automated approach represented a promising step toward real-time, on-demand odor visualization. Taken together, our findings highlight the potential of odor visualization as a design paradigm that can be adapted for diverse applications. We envision further integration of real-time odor sensing, personalized prompt engineering, and culturally aware modules to deepen the multisensory and expressive qualities of \textit{Paint by Odor}.

\section{CRediT authorship contribution statement}
\label{sec10}
\textbf{Gang Yu:} Writing – original draft, Conceptualization, Methodology, Formal analysis, Investigation, Software, Data curation, Visualization. \textbf{Yuchi Sun:} Writing – original draft, Conceptualization, Methodology, Formal analysis, Investigation, Data curation. \textbf{Weining Yan:} Writing – original draft, Formal analysis, Data curation, Software. \textbf{Xinyu Wang:} Data curation, Software. \textbf{Qi Lu:} Writing – review \& editing, Conceptualization, Methodology, Supervision, Resources, Funding acquisition.

\section{Declaration of competing interest}
\label{sec11}
The authors declare that they have no known competing financial interests or personal relationships that could influence the work reported in this paper.

\section{Acknowledgments}
\label{sec12}
This work was supported by the Foundation of National Key Laboratory of Human Factors Engineering (Grant NO. HFNKL2023J08) and the National Natural Science Foundation of China(Grant No. 62441219).
\bibliographystyle{elsarticle-harv}
\bibliography{references}

\appendix

\section{Papers for training the expert GPT}

\setcounter{table}{4}  
\renewcommand{\thefigure}{A.\arabic{table}} 

\begin{table}[h]
\resizebox{\textwidth}{!}{
\begin{tabular}{@{}ll@{}}
    \toprule
    \textbf{Title} & \textbf{Authors}\\
    \midrule
    \makecell[tl]{Visual Aesthetics and Human Preference} & Stephen E et al. \\
    \makecell[tl]{Human electroencephalographic (EEG) response to olfactory stimulation: \\ Two experiments using the aroma of food} & G. Neil Martin \\
    \makecell[tl]{Can Eyes Smell? Cross-Modal Correspondences Between Color Hue-Tone \\ and Fragrance Family} & Yu-Jin Kim \\
    \makecell[tl]{Visualising fragrances through colours: The mediating role of emotions} & \makecell[tl]{Hendrik N J Schifferstein \\ and Inge Tanudjaja} \\
    \makecell[tl]{Making Sense of Scents: The Colour and Texture of Odours} & \makecell[tl]{Ferrinne Spector \\ and Daphne Maurer} \\
    \makecell[tl]{Smelling Shapes: Crossmodal Correspondences Between Odors and Shapes} & Grant Hanson-Vaux et al. \\
    \makecell[tl]{Chocolate smells pink and stripy: Exploring olfactory- visual synesthesia} & Alex Russell \\
    \makecell[tl]{Crossmodal Associations Between Olfaction and Vision: \\ Color and Shape Visualizations of Odors} & Kathrin Kaeppler \\
    \makecell[tl]{Recognition of common odors, pictures, and simple shapes} & HARRY T. LAWLESS \\
    \makecell[tl]{Contribution to understanding odour–colour associations} & Yelena Maric \\
    \makecell[tl]{Olfactory–Visual Congruence Effects Stable Across Ages: \\ Yellow Is Warmer When It Is Pleasantly Lemony} & Estelle Guerdoux et al. \\
    \makecell[tl]{Topographical EEG maps of human responses to odors} & W.R.Klemm et al. \\
    \makecell[tl]{The nose tells it to the eyes: \\ Crossmodal associations between olfaction and vision} & Alix Seigneuric et al. \\
    \makecell[tl]{A Multilingual Benchmark to Capture Olfactory Situations over Time} & S. Menini et al. \\
    \makecell[tl]{When the Sense of Smell Meets Emotion: Anxiety-State-Dependent \\ Olfactory Processing and Neural Circuitry Adaptation} & \makecell[tl]{Elizabeth A. Krusemark \\ et al.} \\
    \makecell[tl]{The importance of the olfactory system in human well‑being, \\ through nutrition and social behavior} & \makecell[tl]{Sanne Boesveldt and \\ Valentina Parma} \\
    \makecell[tl]{Cross-Cultural Color-Odor Associations} & Carmel A. Levitan \\
    \makecell[tl]{Cross-Modal Correspondence between Vision and Olfaction: \\ The Color of Smells} & Avery N. Gilbert et al. \\
    \makecell[tl]{Color–Odor Interactions: A Review and Model} & Debra A. Zellner \\
    \bottomrule
\end{tabular}
}
\caption{Detailed Information of Participants in Formal Study}
\label{tabS2}
\end{table}

\section{Participant Information for Formal Study}

\setcounter{table}{5}  
\renewcommand{\thefigure}{B.\arabic{table}} 

\begin{table}[h]
\begin{tabular}{@{}llll@{}}
    \toprule
    \textbf{Participant} & \textbf{Age} & \textbf{Gender} & \textbf{Education Background} \\ 
    \midrule
    P1 & 19 & M & \makecell[tl]{Broadcast and Television \\ Program Production} \\
    P2 & 20 & M & Robotics and Artificial Intelligence \\
    P3 & 26 & M & Graphic Design \\
    P4 & 33 & M & Computer Science \\
    P5 & 16 & M & None \\
    P6 & 54 & F & None \\
    P7 & 47 & M & Media Technology \\
    P8 & 40 & F & Chemistry \\
    P9 & 29 & F & Chemical Engineering \\
    P10 & 32 & F & Traditional Chinese Medicine \\
    P11 & 17 & M & None \\
    P12 & 31 & M & Industrial Design \\
    P13 & 19 & M & \makecell[tl]{Creative Design and \\ Intelligent Engineering} \\
    P14 & 23 & F & Interaction Design \\
    P15 & 20 & M & \makecell[tl]{Theoretical and Applied Mechanics; \\ Biomedical Engineering} \\
    P16 & 19 & F & Cognitive Science \\
    P17 & 30 & F & Psychology \\
    P18 & 23 & F & Medicine \\
    P19 & 22 & M & Music \\
    P20 & 19 & F & Communication Engineering \\
    P21 & 34 & M & Food Engineering \\
    P22 & 18 & F & Communication Engineering \\
    P23 & 40 & M & In-Car Air Quality \\
    P24 & 28 & F & Architecture \\
    P25 & 26 & F & Education \\
    P26 & 23 & F & Design \\
    P27 & 21 & F & Digital Media Arts \\
    P28 & 36 & F & Fine Arts and Psychology \\
    \bottomrule
\end{tabular}
\caption{Detailed Information of Participants in Formal Study}
\label{tabS1}
\end{table}

\section{Distribution of Questionnaire Answers}

\setcounter{figure}{14}  
\renewcommand{\thefigure}{C.\arabic{figure}} 

\begin{figure}[t]
\centering
\includegraphics[width=0.7\columnwidth]{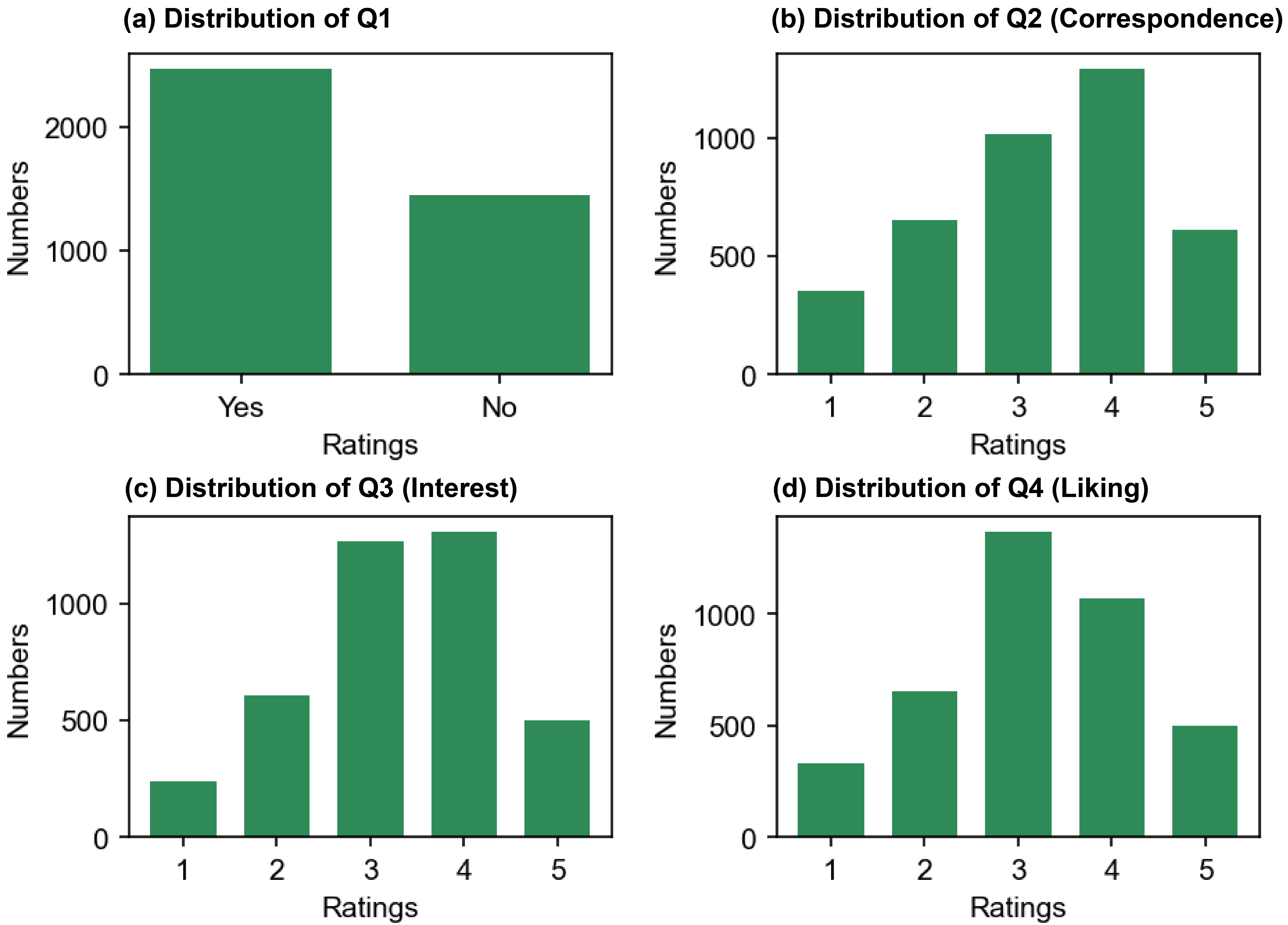}
\caption{Distribution of quantitative data showing non-normal distributions for all four questions.}
\label{fig:norm}
\end{figure}

\section{Variations in GPT's answer}

\setcounter{table}{6}  
\renewcommand{\thefigure}{D.\arabic{table}} 

\begin{table}[h]
\resizebox{\textwidth}{!}{
\begin{tabular}{@{}lll@{}}
    \toprule
    \textbf{Odor Substance} & \textbf{\makecell[cl]{First Query\\(GPT4.0 2024.04)}} & \textbf{\makecell[cl]{Second Query\\(GPT4.0 2024.11)}}\\
    \midrule
    Lemon & Fruity,Fresh,Citrus,Lemon & lemon,fresh,sour \\
    Grape & Fruity,Juicy,Fresh,Grape & fruity,grape \\
    Tea & Earthy,Floral,Woody,Fresh,Dry,Roasted,Mint & fresh,grassy,tea \\
    Baijiu & Winey,Fruity,Earthy,Fermented,Pungent & pungent,fermented,strong \\
    Coffee Bean & Roasted,Warm,Earthy,Caramellic,Coffee,Burnt & roasted,coffee,earthy \\
    Vinegar & Sour,Pungent,Fermented,Acrid & sour,pungent \\
    Sesame Oil & Roasted,Warm,Nutty,Oily & oily,earthy \\
    Cigarette & Tobacco,Earthy,Woody,Leathery & tobacco,earthy \\
    Lily Flower & Floral,Fresh,Sweet & floral,fresh \\
    Latex Paint & Chemical,Pungent,Fresh & chemical,pungent \\
    Varnish & Pungent,Chemical,Woody & waxy,pungent \\
    Mark Pen Ink & Chemical,Pungent,Sharp & medicinal,pungent \\
    Glue & Chemical,Pungent,Sharp & chemical,pungent \\
    Shampoo & Fruity, Floral, Fresh, Mint & floral, fresh, fruity \\
    Shower Gel & Fruity,Floral,Fresh,Mint & floral,fresh,fruity \\
    Detergent & Floral,Fresh,Lemon,Chemical & fresh,fruity,floral \\
    Mouthwash & Mint, Fresh, Medical, Pungent  & minty,fresh,medicinal \\
    Osmanthus Essential Oil & Floral, Fruity, Sweet, Fresh & floral,fruity,fresh \\
    Shoe Polish & Waxy,Oily,Leathery,Earthy & waxy,oily,smoky \\
    Laundry Liquid & Floral,Fresh,Clean & fresh,floral,fruity \\
    \bottomrule
\end{tabular}
}

\caption{The two times of asking LLM choosing odor descriptors.}
\label{tabS3}
\end{table}





\end{document}